\documentclass[twocolumn]{aastex631}

\newcommand{\hii}{H\,{\sc{ii}}}
\newcommand{\hchii}{HC H\,{\sc{ii}}}
\newcommand{\uchii}{UC H\,{\sc{ii}}}
\newcommand{\hmole}{H$_2$}

\newcommand{\msun}{$\rm M_\odot$}

\newcommand{\mjybeam}{mJy~beam$^{-1}$}

\newcommand{\jybeam}{Jy~beam$^{-1}$}

\newcommand{\cmsquare}{cm$^{-2}$}
\newcommand{\cmcube}{cm$^{-3}$}

\newcommand{\kms}{km~s$^{-1}$}

\newcommand{\te}{$T_{\rm e}$}

\newcommand{\nelectron}{$n_{\rm e}$}

\newcommand{\ci}{$c_{\rm i}$}

\newcommand{\htcop}{H$^{13}$CO$^+$}

\newcommand{\htcn}{H$^{13}$CN}

\newcommand{\tcs}{$^{13}$CS}
\newcommand{\sio}{SiO}

\newcommand{\chtoh}{CH$_3$OH}

\newcommand{\so}{SO}

\newcommand{\cthfcn}{C$_2$H$_5$CN}
\newcommand{\chtcn}{CH$_3$CN}
\newcommand{\hfourtyalpha}{H40$\alpha$}
\newcommand{\hthirtyalpha}{H30$\alpha$}

\newcommand{\htcopone}{H$^{13}$CO$^+$\,$J=1-0$}

\newcommand{\siofive}{SiO\,$J=5-4$}

\newcommand{\cthfcnline}{C$_2$H$_5$CN\,$J=26-25$}

\newcommand{\ceo}{C$^{18}$O}

\newcommand{\ceotwo}{C$^{18}$O\,$J=2-1$}

\newcommand{\dustt}{$T_{\rm dust}$}
\newcommand{\vlsr}{$\rm v_{lsr}$}

\usepackage{graphicx} 
\usepackage{rotating}
\usepackage{amsmath} 
\usepackage[rightcaption]{sidecap}

\begin{document}

\title{ALMA-QUARKS: Few-Thousand-Year Hatching out of ``Egg'': The Supersonic Breakout of a Hypercompact \hii\ Region from Its Parental Hot Core}

\author[0000-0002-9836-0279]{Siju Zhang}
\affiliation{Departamento de Astronom\'{i}a, Universidad de Chile, Camino el Observatorio 1515, Las Condes, Santiago, Chile}
\affiliation{Chinese Academy of Sciences South America Center for Astronomy, National Astronomical Observatories, CAS, Beĳing 100101, China}

\author[0000-0003-1649-7958]{Guido Garay}
\affiliation{Departamento de Astronom\'{i}a, Universidad de Chile, Camino el Observatorio 1515, Las Condes, Santiago, Chile}
\affiliation{Chinese Academy of Sciences South America Center for Astronomy, National Astronomical Observatories, CAS, Beĳing 100101, China}

\author[0000-0001-5950-1932]{Fengwei Xu}
\affiliation{Kavli Institute for Astronomy and Astrophysics, Peking University, 5 Yiheyuan Road, Haidian District, Beijing 100871, China}
\affiliation{Max Planck Institute for Astronomy, K{\"o}nigstuhl 17, 69117 Heidelberg, Germany}

\author[0000-0003-2737-5681]{Luis F. Rodr\'{i}guez}
\affiliation{Instituto de Radioastronom\'{\i}a y Astrof\'{\i}sica,
Universidad Nacional Aut\'onoma de M\'exico, Apdo. Postal 3-72, Morelia, Michoac\'an 58089, Mexico}

\author[0000-0001-5175-1777]{Neal J. Evans II}
\affiliation{Department of Astronomy, The University of Texas at Austin, 2515 Speedway, Austin, TX 78712, USA}

\author[0000-0001-9509-7316]{Annie Zavagno}
\affiliation{Aix Marseille Univ, CNRS, CNES, LAM, F-13388 Marseille, France}
\affiliation{Institut Universitaire de France, Paris, 1 rue Descartes, F-75231 Paris Cedex 05, France}

\author[0000-0002-6622-8396]{Paul F. Goldsmith}
\affiliation{Jet Propulsion Laboratory, California Institute of Technology, 4800 Oak Grove Drive, Pasadnea CA 91109, USA}

\author[0009-0004-6159-5375]{Dongting Yang}
\affiliation{School of Physics and Astronomy, Yunnan University, Kunming, 650091, China}

\author[0000-0001-8315-4248]{Xunchuan Liu}
\affiliation{Shanghai Astronomical Observatory, Chinese Academy of Sciences, Shanghai 200030, China}

\author[0000-0003-4546-2623]{Aiyuan Yang}
\affiliation{National Astronomical Observatories, Chinese Academy of Sciences, Beijing 100101, China}
\affiliation{Key Laboratory of Radio Astronomy and Technology, Chinese Academy of Sciences, A20 Datun Road, Chaoyang District, Beijing, 100101, China}

\author[0000-0002-5286-2564]{Tie Liu}
\affiliation{Shanghai Astronomical Observatory, Chinese Academy of Sciences, Shanghai 200030, China}

\author[0000-0003-2300-8200]{Amelia M.\ Stutz}
\affiliation{Departamento de Astronom\'{i}a, Universidad de Concepci\'{o}n,Casilla 160-C, Concepci\'{o}n, Chile}

\author[0000-0003-3343-9645]{Hong-Li Liu}
\affiliation{School of Physics and Astronomy, Yunnan University, Kunming, 650091, China}

\author[0000-0001-9822-7817]{Wenyu Jiao}
\affiliation{Shanghai Astronomical Observatory, Chinese Academy of Sciences, Shanghai 200030, China}

\author[0000-0001-5917-5751]{Anandmayee Tej}
\affiliation{Indian Institute of Space Science and Technology, Thiruvananthapuram 695 547, Kerala, India}

\author{Lei Zhu}
\affiliation{Chinese Academy of Sciences South America Center for Astronomy, National Astronomical Observatories, CAS, Beĳing 100101, China}

\author[0000-0003-2412-7092]{Kee-Tae Kim}
\affiliation{Korea Astronomy and Space Science Institute, 776 Daedeokdae-ro, Yuseong-gu, Daejeon 34055, Republic of Korea}
\affiliation{University of Science and Technology, Korea (UST), 217 Gajeong-ro, Yuseong-gu, Daejeon 34113, Republic of Korea}

\author[0000-0002-8586-6721]{Pablo Garc\'ia}
\affiliation{Chinese Academy of Sciences South America Center for Astronomy, National Astronomical Observatories, CAS, Beĳing 100101, China}
\affiliation{Instituto de Astronom\'ia, Universidad Cat\'olica del Norte, Av. Angamos 0610, Antofagasta, Chile}

\author{Thomas Peters}
\affiliation{Max-Planck-Institut f\"{u}r Astrophysik, Karl-Schwarzschild-Str. 1, D-85748 Garching, Germany}

\author[0000-0002-9277-8025]{Thomas M\"{o}ller}
\affiliation{I. Physikalisches Institut der Universit{\"a}t zu K{\"o}ln, Z{\"u}lpicher Str. 77, 50937, K{\"o}ln, Germany}

\author[0000-0003-1275-5251]{Shanghuo Li}
\affiliation{School of Astronomy and Space Science, Nanjing University, Nanjing 210093}

\author[0000-0002-9574-8454]{Leonardo Bronfman}
\affiliation{Departamento de Astronom\'{i}a, Universidad de Chile, Camino el Observatorio 1515, Las Condes, Santiago, Chile}


\correspondingauthor{Siju Zhang and Fengwei Xu}
\email{sijuzhangastro@gmail.com; fengweilookuper@gmail.com}



\begin{abstract}
The kinematic evolution of hypercompact \hii\ (\hchii) regions  around young high-mass stars remains poorly understood due to complex interactions with parental environs. We present ALMA QUARKS/ATOMS 1.3~mm/3~mm observations (the highest resolution $\sim0.01$~pc) of a deeply embedded \hchii\ region (diameter $\sim0.015$~pc, electron density $\sim2\times10^{5}$~\cmcube) exhibiting a striking $\gtrsim20$~\kms\ global redshift seen in optically thin \hthirtyalpha/\hfourtyalpha\ recombination lines relative to its parental hot molecular core within a hub-filament system. The 1.3~mm continuum data reveal a distinct 0.1-pc arc and a perpendicular 0.04-pc tail. We propose that this morphology arises from a dynamic champagne flow: the slow expansion of \hchii\ region into a pre-existing filament forms the arc and associated low-velocity (few \kms) \sio\ shocks. Meanwhile, in the opposite direction ionized gas escapes along a steep density gradient traced by the tail and high-velocity (20~\kms) \sio\ emission. We reject the bow shock scenario in which ionized gas co-moves with a runaway high-mass star because shocked gas in the arc aligns with the hub velocity, contradicting the bow shock prediction. Non-LTE radiative transfer modeling further rules out infall of ionized gas as the velocity shift origin. We conclude that this exceptional \hchii\ region is undergoing a few-thousand-year transition phase of ``hatching out of the egg'': the ionized gas of \hchii\ region has just broken out of its parental hot core and now is flowing outward supersonically. This work highlights how anisotropic density distributions induce supersonically anisotropic ionized flows that govern \hchii\ region evolution.

\end{abstract}



\section{introduction} \label{sec:introduction}
The formation of high-mass stars is markedly more complex than that of low-mass stars, owing to intense feedback such as photoionization and stellar winds \citep{Beuther2025,Suin2025}. The ultraviolet radiation emitted by nascent high-mass stars ionizes the surrounding neutral medium, generating hypercompact \hii\ (\hchii) regions (electron density $n_{\rm e}\gtrsim10^6~\rm cm^{-3}$ and diameter $D\lesssim0.03~\rm pc$) that suppress sustained accretion and evolve through phases of expansion, eventually transitioning to classical \hii\ regions \citep{Kurtz2005}. Investigating the kinematics of these nascent \hchii\ regions is therefore critical to elucidate the early stages of high-mass star formation (HMSF). Observations and numerical simulations have revealed intricate gas kinematic processes in \hchii\ regions, including molecular/ionized disk-jet systems \citep{Guzman2020,Moscadelli2021, Miyawaki2023}, accretion/infall in ionized and molecular flows \citep{Keto2003, Peters2010a,Peters2010b, Komesh2024}, and highly variable expansion or even contraction \citep{Galvan-Madrid2008, Peters2010a, DePree2014,Yang2025}.

The kinematics of \hii\ regions is intrinsically linked to their parental molecular clumps, often inheriting their systemic velocities from the precursor clumps \citep{Li2023}. However, this inheritance correlation breaks down if the ionizing sources are runaway stars originating externally or if the ionized gas breaks out and escapes from the natal clump. Depending on stellar motion within the natal clump, ionized gas may co-move with the ionizing star to form bow shocks \citep{vanBuren1990} or flow along density gradients as champagne flows \citep{Tenorio1979}. Previous observations of such dynamical systems frequently focused on ultracompact \hii\ (\uchii) regions ($n_{\rm e}\gtrsim10^4~\rm cm^{-3}$ and $D\lesssim0.1~\rm pc$) that exhibit cometary morphologies \citep[][and references therein]{Garay1999, Kim2001,Kim2003,Churchwell2002,Hoare2005,Hoare2007,Zhang2024}, due to the scarcity and compactness of \hchii\ regions \citep{Yang2019,Yang2021}. Furthermore, high-frequency radio recombination lines (RRLs) at mm wavelength are essential to characterize the kinematics of ionized gas at extremely high density because they suffer less from pressure broadening, non-LTE effects, and ensure optically thin conditions \citep{Keto2003, Peters2012}.

\begin{figure}[htb!]
\centering
\includegraphics[width=0.45\textwidth]{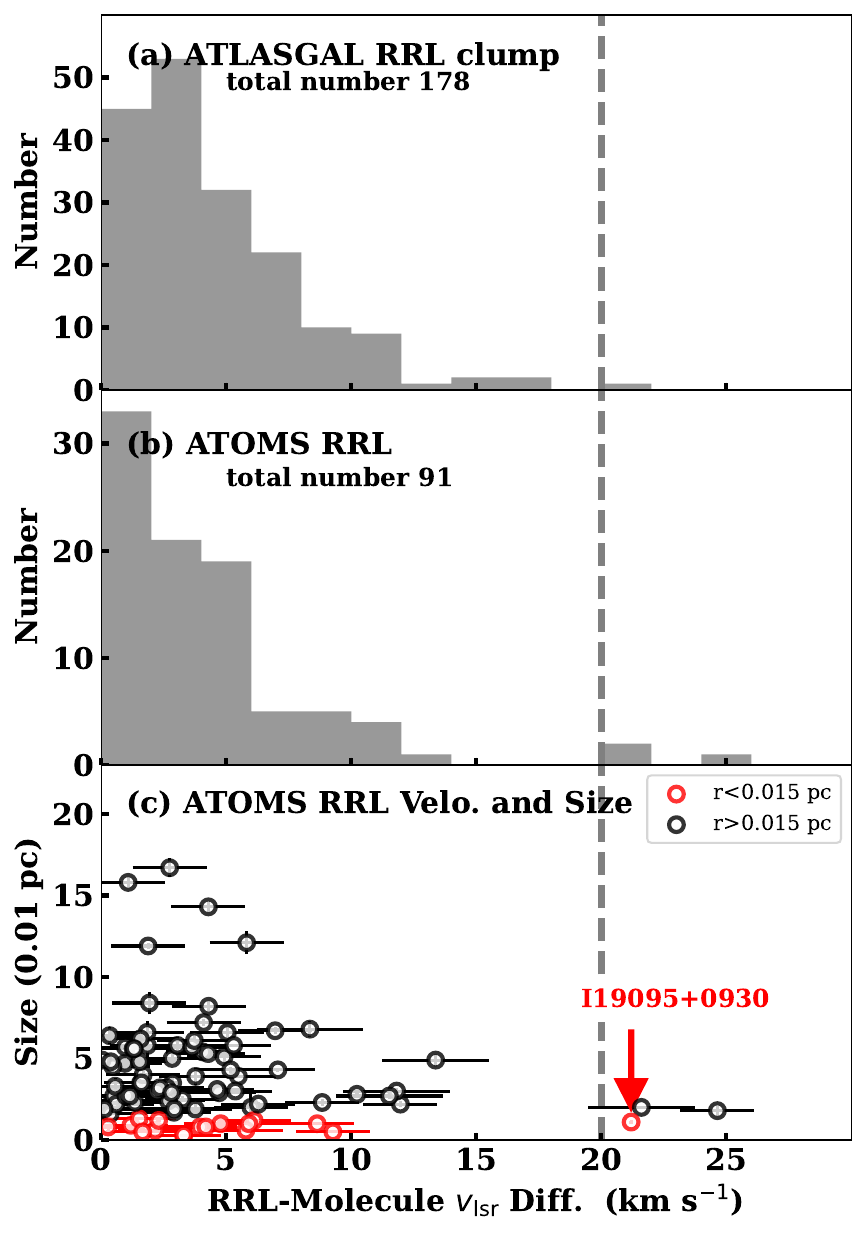}
       \caption{Velocity shifts between mm RRLs and molecular lines in high-mass star-forming molecular clumps. Vertical dashed line highlights the shift of 20~\kms. (a) Single-dish survey of 976 ATLASGAL clumps \citep[RRLs detected in 178 sources;][]{Kim2017}. (b) Velocity shifts for 91 embedded HC and \uchii\ regions, identified across 146 ATOMS molecular clumps \citep{ATOMSIII}. (c) Size (radius)-velocity shift relation for 91 ATOMS HC or \uchii\ regions, highlighting compactness and extreme kinematics of I19095.}
        \label{fig:velo-comparison}
\end{figure}

To trace the dynamical decoupling between the ionized gas and the parental molecular clump, we analyzed the offsets of systemic velocity measured with mm RRLs and molecular lines in two surveys, shown in Fig.~\ref{fig:velo-comparison}. Panel (a) shows the parsec-scale velocity offsets in single-dish surveys of ATLASGAL clumps \citep{Kim2017} and panel (b) shows the ATOMS \citep[ALMA Three-millimeter Observations of Massive Star-forming regions, resolution $\sim1.5$\arcsec;][]{ATOMSI, ATOMSIII} observations of UC and \hchii\ regions still embedded in molecular clumps. Line tracers are the same within each survey. From Figure~\ref{fig:velo-comparison}, we can see that extreme velocity shifts ($>20$~\kms) are very rare ($<2\%$) among the detected clumps and cores. In particular, the \hchii\ region in IRAS 19095+0930 \citep[I19095 hereafter;][]{ATOMSIII} exhibits the most extreme velocity shift and compactness within the sample. The other two \hii\ regions with an extreme velocity shift ($>20$~\kms) are found to have explosive outflows \citep{Zapata2023} or locate in the Central Molecular Zone (thus hard to determine clump velocity). 

Here, we present I19095 ATOMS and high-resolution ($\sim0.3$\arcsec) ALMA-QUARKS observations \citep[Querying Underlying mechanisms of massive star formation with ALMA-Resolved gas Kinematics and Structures;][]{QUARKSI}. Our analysis reveals that the kinematics of this \hchii\ region can be interpreted as a champagne flow associated with a hot core and defies interpretation in terms of bow shocks or infalling ionized gas under non-LTE conditions. Our observations of this source challenge the assumption of spherically symmetric expansion even for the earliest \hii\ regions and invoke a very rapid transition phase of \textit{``hatching out of the egg''} in the hot core-\hchii\ region evolution.

\section{ALMA Data} \label{sec:data}
QUARKS 1.3~mm data combine observations from the 12~m main array (configurations C-2 and C-5) and the Atacama Compact Array (ACA) 7~m array to achieve an angular resolution of $0.35$\arcsec\ and maximum recoverable scale (MRS) of $\sim30$\arcsec. The continuum sensitivity reaches $\sim0.12$~\mjybeam, with spectral line data binned at a channel width of 0.63~\kms\ and a sensitivity of $\sim0.4$~K. More details on the QUARKS dataset are provided in \citet{QUARKSI, QUARKSII,QUARKSIII}.

ATOMS 3~mm data combine the observations in the C-3 configuration and the ACA 7m array. The 12~m and 7~m data sets were subsequently combined and imaged jointly using \texttt{CASA} 6.6.0 \citep{CASA2022}. The final continuum image achieves a sensitivity of 0.13~\mjybeam, with phase solutions derived from the continuum applied to the spectral cubes. The angular resolution and MRS are 1.58\arcsec\ and $\sim60$\arcsec, respectively, with spectral channels of 1.48~\kms\ width and a sensitivity of 0.11~K for \hfourtyalpha\ and 0.21~\kms\ and 0.3~K for \htcopone\ \citep{ATOMSI}.

To improve image fidelity and dynamic range, three rounds of phase self-calibration were applied to both the QUARKS and ATOMS data, effectively reducing sidelobes caused by phase errors. 

\section{Parental hub-filament system} \label{sec:source}
The region I19095, also known as the hydroxyl maser source OH 43.8$-$0.1 \citep{Winnberg1975}, is a highly active site of HMSF, evidenced by the detection of numerous molecular masers. These include OH masers at 1.66~GHz \citep{Fish2005} and 6.03~GHz \citep{Baudry1997}, H$_2$O masers at 22~GHz \citep{Honma2005}, and \chtoh\ Class I \citep[44~GHz;][]{Kim2019} and Class II \citep[6.7~GHz;][]{Surcis2019} masers. Most of the masers are concentrated near the systemic velocity of the local standard of rest ${\rm v_{lsr}}\sim43$~\kms, spanning a range of a few \kms. Trigonometric parallax measurements of H$_2$O masers yield a distance of $6.02^{\rm +0.39}_{\rm -0.34}$~kpc \citep{Wu2019}. At this distance, the natal clump traced by the ATLASGAL 870~\micron\ emission has a mass of $\sim530$~\msun, a radius of 0.24~pc, with a dust temperature \dustt\ of 35~K \citep{Urquhart2022}.

\subsection{Velocity-coherent hub-filament system} \label{subsec:parental-hfs}
\begin{figure}
\centering
\includegraphics[width=0.45\textwidth]{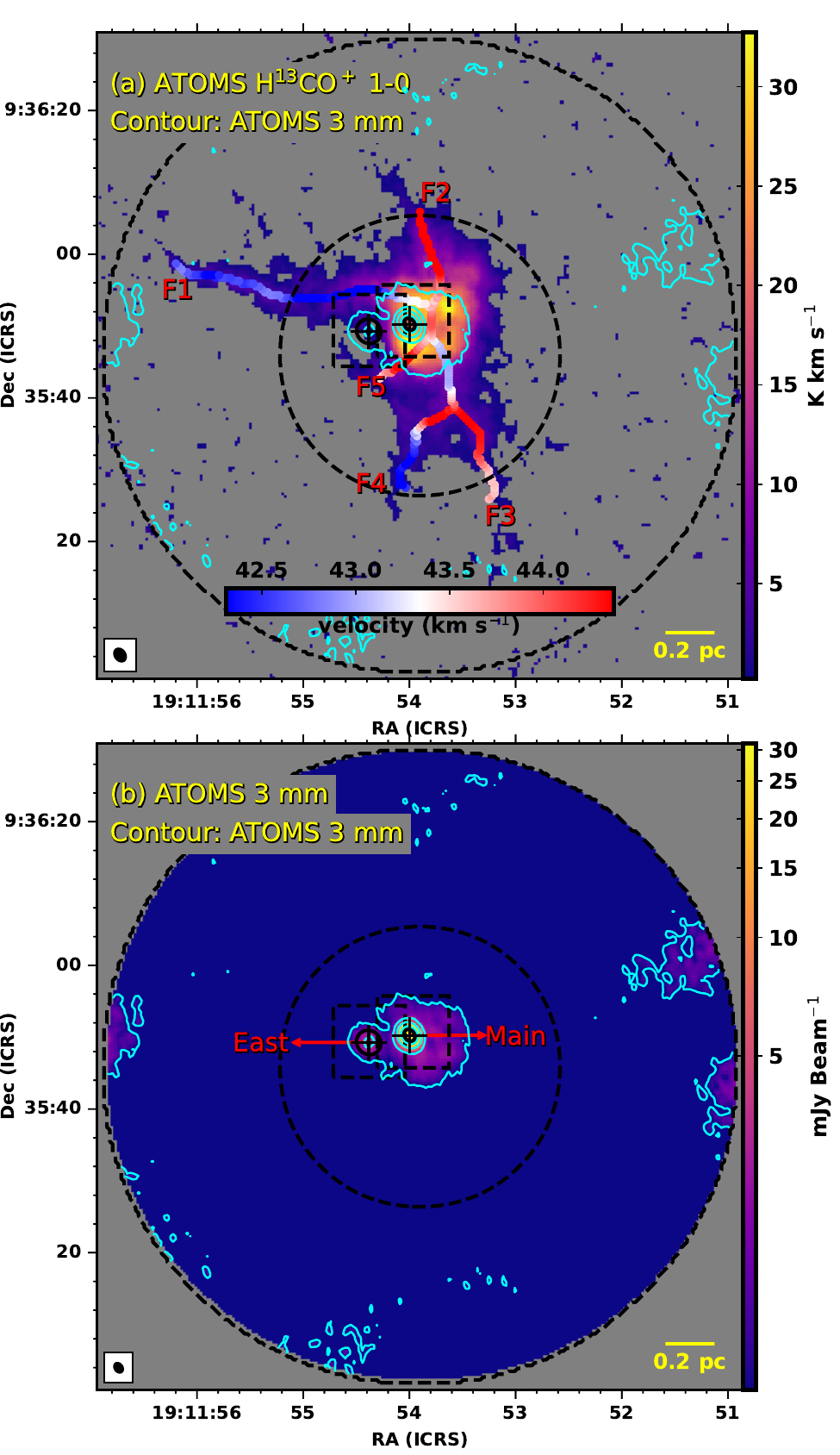}
       \caption{ATOMS view of the I19095 hub-filament system and the two embedded 3 mm cores. (a) ATOMS \htcopone\ 0th moment map ($3\sigma$ channel threshold), overlaid with ATOMS 3~mm contours (levels: 0.72–222~\mjybeam\ in 5 logarithmic steps) and filament skeletons (F1–F5) color-coded by their \vlsr. (b) ATOMS 3~mm continuum emission, with ellipses denoting Gaussian FWHM sizes for the cores, crosses making the core centers, boxes indicating zoom-in regions shown in Fig.~\ref{FIGURE:ATOMS-HII}, and circles marking the fields of view of ATOMS and QUARKS.}
        \label{FIGURE:GLOBAL}
\end{figure}

Figure~\ref{FIGURE:GLOBAL}a displays the 0th moment map integrated from ${\rm v}_{\rm lsr}=35$ to 55~\kms\ of ATOMS \htcopone\ emission that well traces the dense gas in the HMSF regions \citep{Zhang2024}. At least five filamentary structures (F1–F5), identified using the \texttt{FilFinder} algorithm \citep{Koch2015}, converge to the central hub. The filament skeletons, color-coded by \vlsr\ of \htcop, reveal that the entire hub-filament system (HFS) is velocity-coherent, centered at $\rm v_{lsr}\simeq43$~\kms. Notably, some filaments (e.g., F3 and F4) exhibit significant velocity gradients toward the hub, while others show non-monotonic gradients, suggesting complex 3D projection effects if sustained inflow occurs. The filament with the most prominent velocity gradient, F4, shows a projected velocity gradient of $\gtrsim1~\rm km~s^{-1}~pc^{-1}$. Assuming an \hmole\ column density of $\sim10^{23}$~\cmsquare\ derived from \htcop\ using the methods in \citet{Zhang2024}, this gradient implies an inflow rate of $\sim0.1~\rm M_{\odot}~kyr^{-1}$. The total ongoing mass accretion into the hub, if all converging filaments are counted, is therefore likely on the order of 0.1 M$_\odot$ kyr$^{-1}$.

\subsection{3-mm cores and spatially correlated \texorpdfstring{\hfourtyalpha}{}} \label{subsec:3mmcore}
\begin{figure*}
\centering
\includegraphics[width=0.95\textwidth]{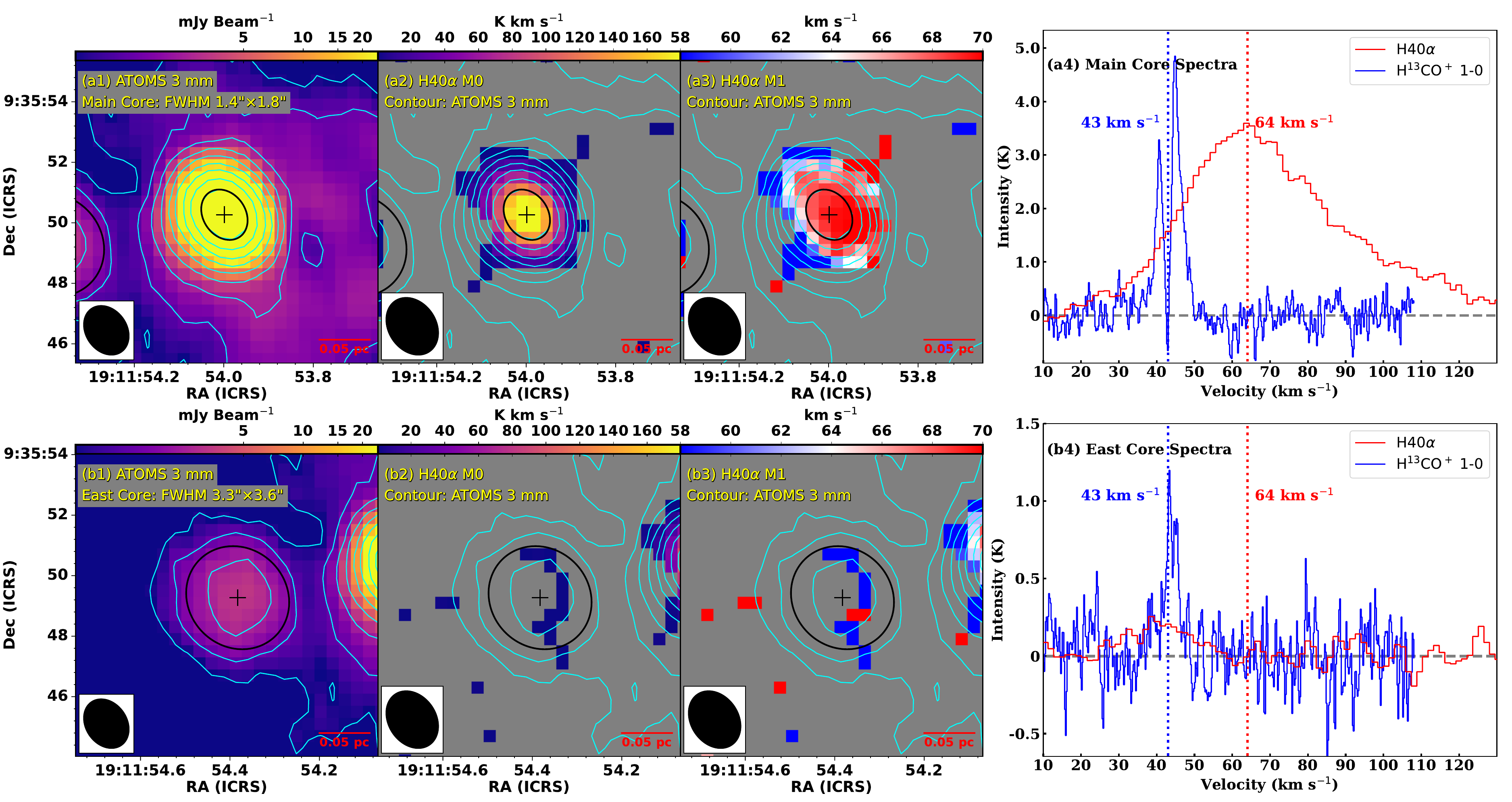}
       \caption{ATOMS 3~mm continuum and \hfourtyalpha\ line emission of the Main Core (first row) and the East Core (second row). (a1) 3~mm continuum. (a2) \hfourtyalpha\ 0th moment map (a3) \hfourtyalpha\ velocity field (1st moment). (a4) Source-averaged \hfourtyalpha\ (red) and \htcop\ (blue) spectra. Panels (b1)–(b4) follow the same order as (a1)–(a4). Contours (levels: 0.72–222~\mjybeam\ in 10 logarithmic steps) trace the 3-mm continuum emission. Ellipses denote Gaussian FWHM sizes. All moment maps are generated using $3\sigma$ channel threshold.}
        \label{FIGURE:ATOMS-HII}
\end{figure*}

The ATOMS 3~mm continuum image (Fig.~\ref{FIGURE:GLOBAL}b) reveals two compact sources associated with the ATOMS \hfourtyalpha\ line emission which are shown in Fig.~\ref{FIGURE:ATOMS-HII}: \textit{Main Core} (located within the hub) and \textit{East Core} (offsets from the hub). Both cores are associated with known \hii\ regions detected in earlier VLA centimeter-wavelength observations \citep{Kurtz1994, Kalcheva2018} and the two cores have a separation of $\sim5.7$\arcsec\ (equivalent to 0.17~pc). The Main Core remains unresolved in the ATOMS continuum image. A Gaussian fit using \texttt{CASA imfit} fails to yield a deconvolved size and returns only a beam-convolved FWHM 1.8\arcsec$\times$1.4\arcsec\ ($0.05~{\rm pc}\times0.04~{\rm pc}$) with a flux density of $224\pm7.6$~mJy. The East Core, previously undetected in the ATOMS survey, is clearly detected in the present study due to the improved data reduction \citep{ATOMSIII}.

The strong \hfourtyalpha\ suggests that free-free emission dominates the 3~mm continuum emission of the Main Core. The simplified expression for the free–free continuum–to–RRL ratio derived by \citet{Garay1986}, under the assumption of optically thin continuum emission (Sect.~\ref{subsec:mechanism}), is
\begin{equation*}
    T_{\rm C}/T_{\rm RRL} = 4.3\times10^{-5}\,T_{\rm e}^{1.15}\Delta \nu_{\rm RRL}\, \nu_{\rm RRL}^{-2.1},
\end{equation*}
where $\nu_{\rm RRL}$ and $\Delta \nu_{\rm RRL}$ denote the RRL frequency (in GHz) and linewidth (in kHz), respectively.
Applying this relation, the free–free contribution inferred from \hfourtyalpha\ accounts for roughly 70\% of the 3 mm flux density, assuming an electron temperature 
\te\ of $10^4$~K as reported by \citet{Zhang2023}.  The Main Core shows that its \hfourtyalpha\ has a $\gtrsim20$~\kms\ redshift relative to the HFS systemic velocity traced by \htcopone\ ($\rm v_{lsr}\sim43$~\kms, Fig.~\ref{FIGURE:ATOMS-HII}a4). In contrast, such an offset is not detected in the East Core, although its \hfourtyalpha\ is marginally detected ($2\sigma$, see Fig.~\ref{FIGURE:ATOMS-HII}b4).

The unresolved nature of Main Core in ATOMS data motivates the use of higher-resolution QUARKS data to unravel its substructures and kinematics.

\section{ionized gas of the \texorpdfstring{\hchii}{} region} \label{sec:hchii-region}
\begin{figure*}
\centering
\includegraphics[width=0.9\textwidth]{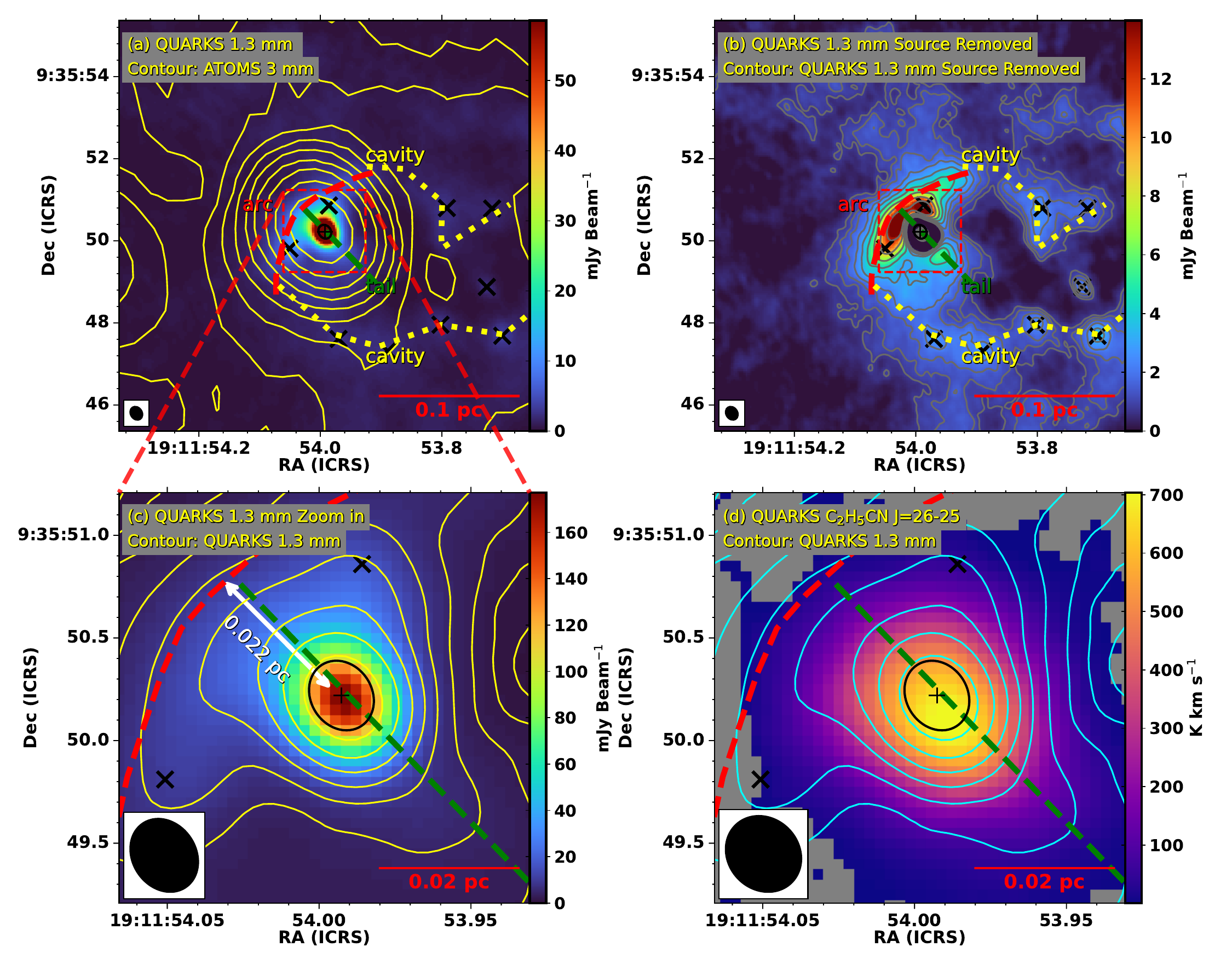}
       \caption{QUARKS images of Main Core and its associated structures. (a) QUARKS 1.3~mm continuum overlaid with ATOMS 3~mm contours (levels identical to Fig.~\ref{FIGURE:ATOMS-HII}) and other extracted 1.3~mm cores (crosses; Jiao et al. in prep.). The extended dusty structures are highlighted with red, green and yellow dashed/dotted lines. (b) QUARKS 1.3~mm continuum after subtracting the \texttt{CASA}-fitted Main Core emission. Associated contour levels are from $6\sigma$ to 17.9~\mjybeam\ in 10 logarithmically spaced steps. (c) Zoomed-in view of Main Core in 1.3~mm continuum with contours (levels: $6\sigma$ to 0.18~\jybeam\ in 10 logarithmically spaced steps). The separation between Main Core and arc vertex is indicated with white scalebar. (d) 0th moment map of \cthfcnline\ overlaid with 1.3~mm contours, tracing hot molecular gas.}
        \label{FIGURE:QUARKS-HII-EXTEND}
\end{figure*}

Figure~\ref{FIGURE:QUARKS-HII-EXTEND}a shows the QUARKS 1.3 mm continuum emission associated with the Main Core region, while Figure~\ref{FIGURE:QUARKS-HII-EXTEND}b presents the image after subtracting the \texttt{CASA}-fitted Main Core 1.3 mm component. This subtraction alleviates the contrast issue and clearly reveals extended substructures related to the Main Core, including three prominent morphological features: the arc, the tail, and the cavity. The Main Core has a 1.3~mm integrated flux density of $306\pm22~\rm mJy$, a beam-convolved FWHM of 0.52\arcsec$\times$0.44\arcsec, and a deconvolved FWHM of 0.36\arcsec$\times$0.3\arcsec\ (corresponding to 2170~au $\times$ 1990~au). The 1.3~mm continuum is spatially coincident with \hthirtyalpha\ and the hot core dense gas tracer \cthfcnline\ \citep[234.42396~GHz, $E_{\rm u} > 100~\rm K$;][]{QUARKSI} line emission in the QUARKS data as shown by  Figs.~\ref{FIGURE:QUARKS-HII-EXTEND}c, d and \ref{FIGURE:QUARKS-HII-GAS}, confirming that a \hchii\ region resides within the hot molecular core--a deeply embedded molecular environment.

\subsection{\texorpdfstring{\hthirtyalpha}{} reconstructed from blended line profile} \label{subsec:h30axclass}
Owing to the deeply embedded nature of \hchii\ region, \hthirtyalpha\ (231.900928~GHz) is spectrally blended with several narrow lines of molecules at similar frequencies, including HNCO, CH$_3$OCH$_3$, C$_2$H$_5$OH, CH$_3$OCHO, and CH$_3^{18}$OH. This makes it complicated to accurately derive the spatial distribution and kinematics of the ionized gas. To address this, we employ the eXtended \texttt{CASA} Line Analysis Software Suite \citep[\texttt{XCLASS};][]{XCLASS2017} for pixel-by-pixel spectral deblending. \texttt{XCLASS} models molecular and recombination lines, by solving the 1D radiative transfer equation under LTE and isothermal assumptions for \hthirtyalpha. The details of the deblending process are described in Appendix~\ref{APP:XCLASS}.

Panels (a-e) of Fig.~\ref{FIGURE:QUARKS-HII-GAS} display the \texttt{XCLASS}-deblended \hthirtyalpha\ maps of electron emission measure (EM), centroid velocity, line width, the source-averaged \hthirtyalpha\ and some other molecular line profiles, respectively. The EM map closely mirrors the morphology of the 1.3~mm continuum emission, with its major axis likely following NE-SW direction (Fig.~\ref{FIGURE:QUARKS-HII-GAS}a). A global velocity gradient from $\rm v_{lsr}\approx64$~\kms\ (NE) to $\rm v_{lsr}\gtrsim66$~\kms\ (SW) is \textit{probably} present in the deblended \hthirtyalpha\ centroid velocity map (Fig.~\ref{FIGURE:QUARKS-HII-GAS}b). A similar but more pronounced gradient is seen in the \hfourtyalpha\ first-moment map, which is redshifted to $\rm v_{lsr}\approx70$~\kms\ (Fig.~\ref{FIGURE:ATOMS-HII}a3). Both  \hthirtyalpha\ and \hfourtyalpha\ exhibit positive skewness in their line profiles, indicating stronger red wings (Figs.~\ref{FIGURE:ATOMS-HII}a4 and \ref{FIGURE:QUARKS-HII-GAS}d). The \hfourtyalpha\ peak velocity becomes increasingly redshifted with larger positive skewness along the NE–SW direction (Appendix~\ref{APP:H40aProfile}). If the \hfourtyalpha\ emission represents a smoothed imprint of the higher-resolution \hthirtyalpha\ emission, the true velocity gradient traced by \hthirtyalpha\ is expected to be even more significant than shown in Fig.~\ref{FIGURE:QUARKS-HII-GAS}b. The \texttt{XCLASS} deblending assumes a symmetric (Gaussian + Lorentzian) line profile, and therefore asymmetric line wing emission is partly suppressed, leading to an underestimation of the actual velocity gradient. The asymmetric line profiles, as indicated by the positive skewness of \hthirtyalpha\ and \hfourtyalpha, suggest that the ionized flows are anisotropic and deviate from spherical symmetry if both lines are optically thin \citep{Sewilo2004}. Some QUARKS molecular lines (\ceotwo, \cthfcnline, \siofive) also exhibit positive skewness (Fig.~\ref{FIGURE:QUARKS-HII-GAS}e), with \sio\ and \cthfcn\ can extend up to almost $\rm v_{lsr}\sim60$ and 50~\kms, respectively.  The line width of \hthirtyalpha\ (Fig.~\ref{FIGURE:QUARKS-HII-GAS}c) ranges from 44 to 50~\kms, consistent with the value of typical \hchii\ regions \citep{Keto2008}.

\subsection{Properties of the \texorpdfstring{\hchii}{} region} \label{subsec:hchii}
\begin{figure*}
\centering
\includegraphics[width=0.9\textwidth]{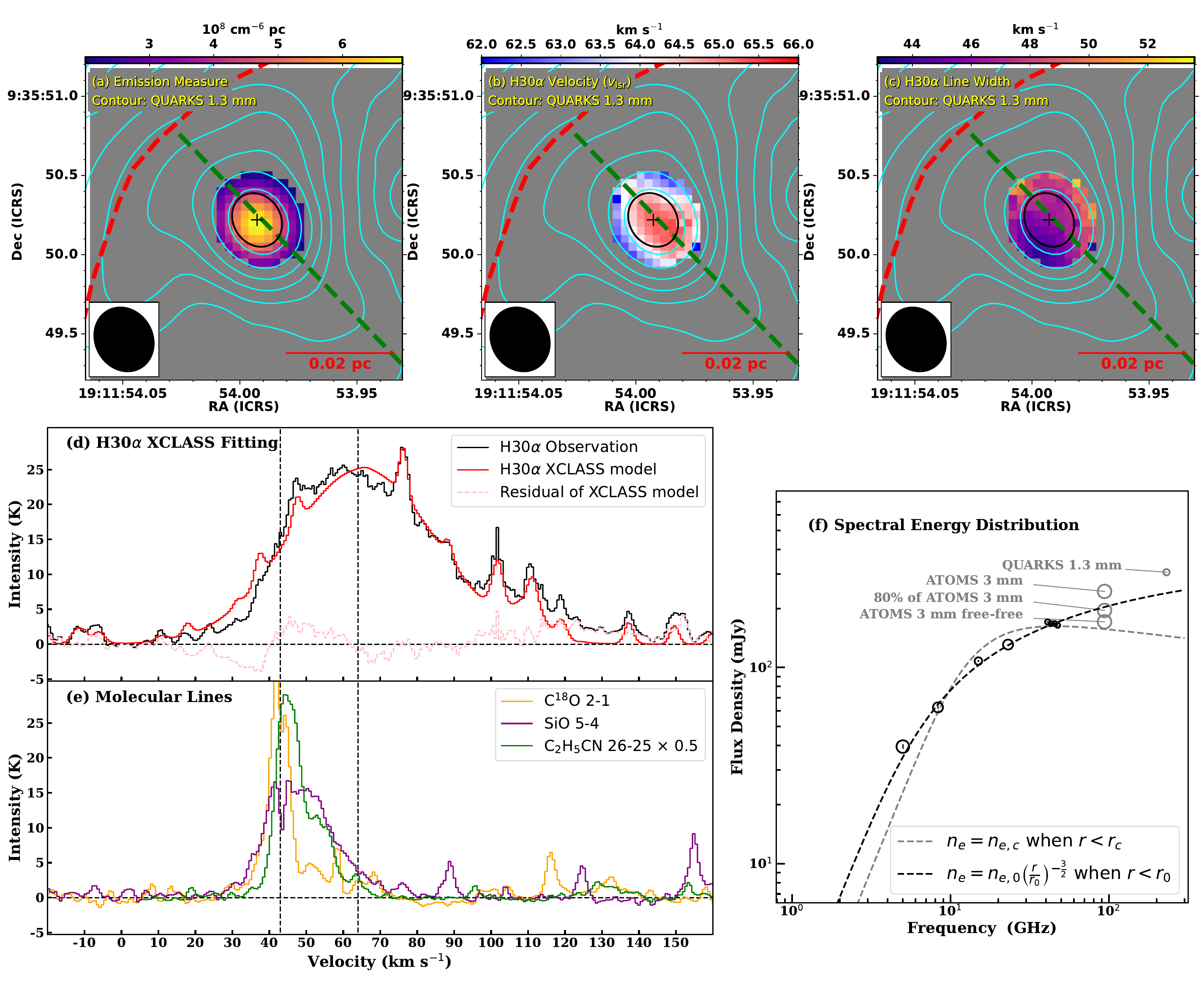}
       \caption{\texttt{XCLASS}-deblended \hthirtyalpha. (a) EM, (b) centroid velocity, and (c) line width. (d, e) source-averaged spectra: observed (black) and modeled (red) \hthirtyalpha, compared with \ceo, C$_{2}$H$_{5}$CN, and SiO. The observed and modeled \hthirtyalpha\ spectra both include the molecular line emission. (f) SED fitted with uniform (gray) and power-law (black) density models. Archival VLA data (7~mm–6~cm; Appendix~\ref{APP:SED}) and ATOMS 3~mm/QUARKS 1.3~mm data are shown as black and gray dots, respectively, with sizes indicating the beam size of observations.}
        \label{FIGURE:QUARKS-HII-GAS}
\end{figure*}

From the deblended \hthirtyalpha\ EM maps in Fig.~\ref{FIGURE:QUARKS-HII-GAS}a, the electron density \nelectron\ is estimated as $\sqrt{EM/2R}\sim 2.0\times10^5$~\cmcube\ assuming uniform density and FWHM radius $R$ measured from EM map (deconvolved, 0.27\arcsec$\times$0.25\arcsec, equivalent to 1560~au). The ionization rate $\dot{N_{\rm i}}=n_{\rm e}^2\alpha_{\rm B} V\sim 5.4\times10^{47}~\rm s^{-1}$, where $\alpha_{\rm B}$ and $V$ are the Case B total recombination coefficient and \hchii\ region volume, respectively \citep{Zhang2024}. The derived $\dot{N_{\rm i}}$ corresponds to an ionizing star of type B0.5 to B1 with a mass of $m_{*}\sim18$~\msun\ \citep{Vacca1996, Keto2003}.

The derived \nelectron\ is about an order of magnitude lower than that of typical \hchii\ regions \citep{Kurtz2005}. \citet{RiveraSoto2020} identified a number of \hii\ regions in W51A with sizes smaller than 0.01 pc and \nelectron\ of $10^4$–$10^5$~\cmcube. They suggested that most of these sources are essentially smaller versions of expanding \uchii\ regions ionized by early B-type stars, as they follow the extrapolated size–\nelectron\ relation established for UC and compact \hii\ regions \citep{Garay1999, Kim2001}. The I19095 \hchii\ region shows a similar size, \nelectron, and ionizing spectral type to those reported by \citet{RiveraSoto2020}, even though it remains deeply embedded within a hot core. Therefore, I19095 likely belongs to the similar class of objects defined by \citet{RiveraSoto2020}, representing a slightly more evolved stage than typical \hchii\ regions.

To further constrain the properties of \hchii\ region, we tested SED models dominated by free-free emission (Fig.~\ref{FIGURE:QUARKS-HII-GAS}f) against the archival VLA data spanning 7~mm to 6~cm \citep[see Appendix~\ref{APP:SED};][]{Kurtz1994,Kavak2021} plus the free-free contribution of ATOMS 3~mm. Two density profiles were tested: 1) Uniform model: $n_{\rm e} = n_{\rm e, c}~\left(r<r_{\rm c}\right)$, yielding $n_{\rm e, c} = 3.2\times10^5$~\cmcube\ and $r_{\rm c} = 1240$~au. 2) Power-law model \citep{Keto2003}: $n_{\rm e} = n_{\rm e, 0} \left(r/r_{\rm 0}\right)^{-3/2}~ \left(r<r_{0}\right)$, yielding $n_{\rm e, 0} = 0.7\times10^5$~\cmcube\ and $r_{\rm 0} = 1670$~au. Both cases align well with the properties derived from RRL, within a factor of 2. The power-law model predicts a 3~mm flux density consistent with the observations, accounting for approximately 80\% of the total 3~mm emission—compatible with the $\sim70\%$ inferred from the  \hfourtyalpha\ line. We tested fittings that excluded the 3~mm free–free component and found that the results remain consistent within the uncertainties. Overall, the power-law model provides a better fit than the uniform-density model, suggesting the presence of an electron density gradient within the \hchii\ region. However, the actual density profile can deviate from the assumed index of $-3/2$, owing to possible departures from spherical symmetry and the limitations of modeling the innermost region \citep{Keto2003}.

The QUARKS 1.3~mm continuum were excluded from the SED analysis due to the heavy contributions from dust emission. After subtracting free-free component in 1.3~mm flux density using the power-law model, the residual 1.3~mm flux density of 162~mJy corresponds to a molecular gas mass of 12~\msun\ and a surface density of 8.9~g~\cmsquare\, assuming a \dustt\ of 160~K from \citet{ATOMSVIII}, a gas-to-dust ratio of 100 and a dust opacity of 0.9~$\rm cm^2~g^{-1}$  \citep{Ossenkopf1994}.


\section{Nature of the supersonic ionized gas}\label{sec:origin}
The observed velocity shift between the ionized and molecular components is well above the sound speed of the ionized gas $c_{\rm i}\sim 10$~\kms\ \citep{Zhang2024} and reveals an extremely dynamic system. These spatially correlated components should exhibit distinct interaction signatures. We first describe the extended structures resolved by 1.3~mm in Sect.~\ref{subsec:extendedstructure} because they can trace the interaction between the ionized gas and the molecular gas and then explore the possible mechanisms driving the supersonic ionized gas in Sect.~\ref{subsec:mechanism}.


\begin{figure*}
\centering
\includegraphics[width=0.97\textwidth]{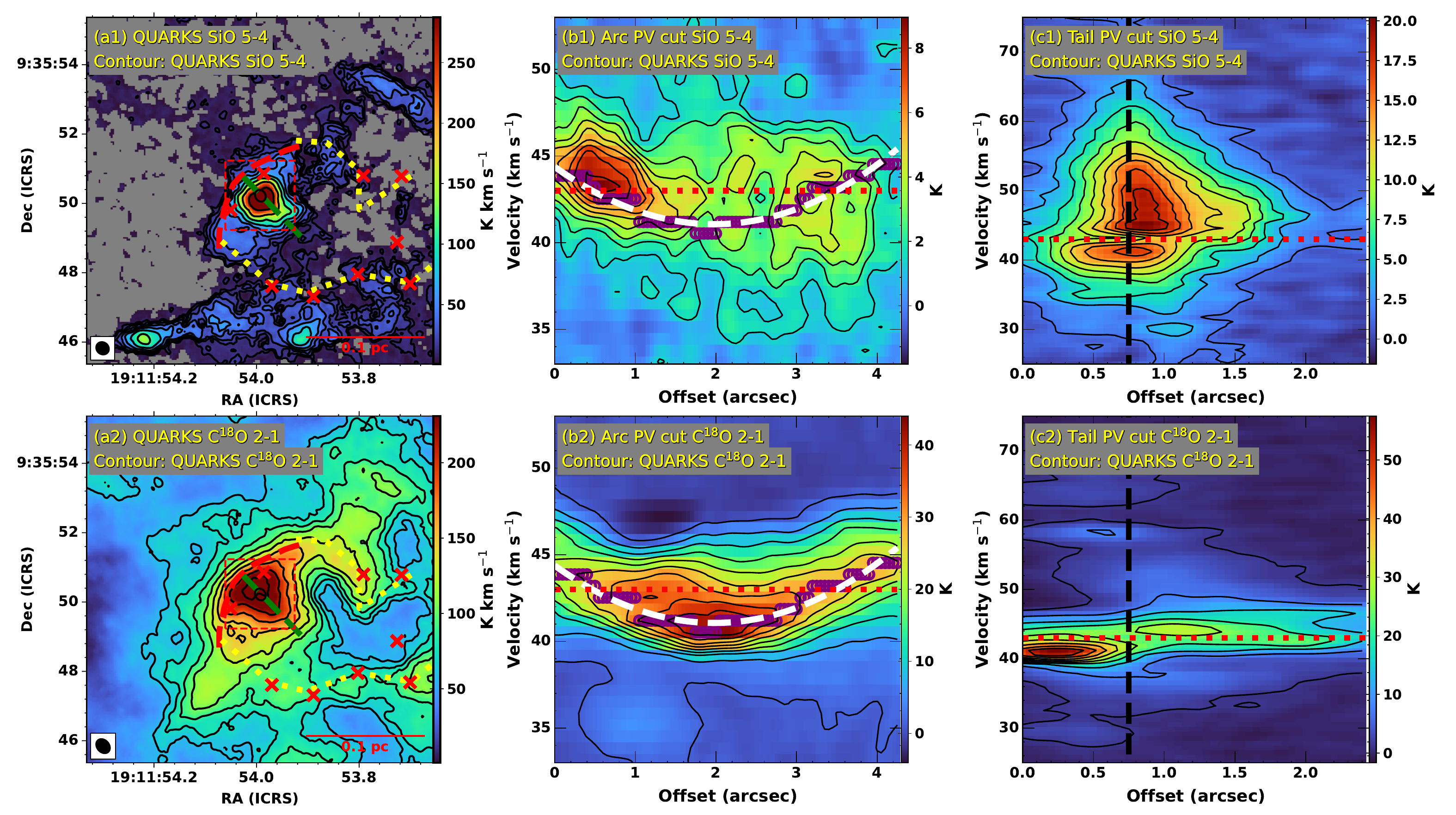} 
\caption{Molecular gas of extended structures. (a1, a2) 0th moment maps with $3\sigma$ channel cutoff of \siofive\ and \ceotwo\ emission, overlaid with 10 logarithmic contours from 9 to 432~$\rm K~km~s^{-1}$ for \sio\ and from 60 to 306~$\rm K~km~s^{-1}$ for \ceo. (b1, b2, c1 and c2) PV diagrams of \sio\ and \ceo\ along arc and tail, respectively (contour levels: 3$\sigma$ to 8.9~K $[b1]$, 44~K $[b2]$, 20.2~K $[c1]$ and 57.5~K$[c2]$ in 10 linear steps). Dotted horizontal lines indicate the systemic velocity of the hub whereas the vertical dashed line in panels c1 and c2 indicate the center position of \hchii\ region. Purple dots and dashed parabolic curve in panels b1 and b2 mark the peak \ceo\ velocity positions and their best-fit parabola.}
        \label{FIGURE:SiOC18OView}
\end{figure*}

\subsection{Extended structures interacting with \texorpdfstring{\hchii}{} region}\label{subsec:extendedstructure}
\textit{The Arc:} An arc structure that extends $\sim0.1~\rm pc$ is visible in the 1.3~mm continuum , \ceotwo, and \siofive\ emission, indicated as red dashed lines in both Figs.~\ref{FIGURE:QUARKS-HII-EXTEND}b and \ref{FIGURE:SiOC18OView}. The continuum-traced arc appears slightly further from the \hchii\ region and encloses the arc traced by \ceo\ and \sio. The arc vertex lies 0.022~pc (0.75\arcsec\,$\pm$\,0.1\arcsec) from the center of the \hchii\ region. The arc mass is around 30 to 90~\msun, estimated with a 1.3~mm flux density of 0.11~Jy and a \dustt\ of 50 to 20~K. Position-velocity diagrams (PVDs) along the arc for \ceo\ and \sio\ are shown in Fig.~\ref{FIGURE:SiOC18OView}. The \ceo\ PVD displays a parabolic shape, also evident in several QUARKS molecular lines (\so, \tcs, \chtoh, \chtcn, and \htcn; see Appendix~\ref{APP:PVDs}). Fitting a parabola to the velocity of the emission peak in each position for \ceo\ PVD yields a vertex velocity of $\sim40$~\kms\ and end velocities between 43 and 45~\kms. In contrast, \sio\ does not clearly show this characteristic of a parabolic shape velocity, but rather exhibits a more consistent velocity around 43~\kms.

\begin{figure*}[htb!]
\centering
\includegraphics[width=0.97\textwidth]{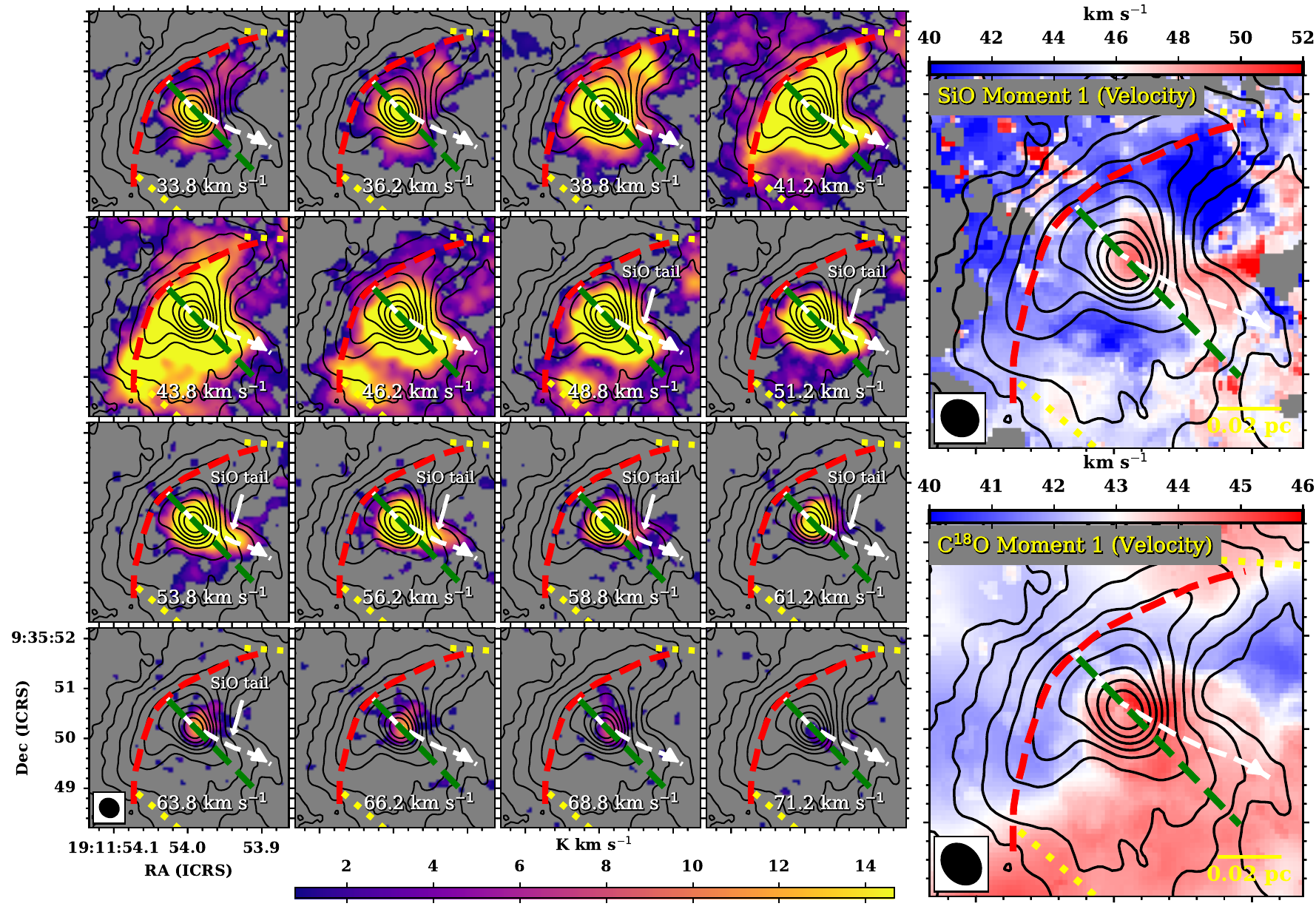}
       \caption{Channel map of \siofive\ (left panels) and the first order moment maps of \siofive\ and \ceotwo\ (right panels). Black contours show 1.3~mm continuum emission, with levels the same as Fig.~\ref{FIGURE:QUARKS-HII-EXTEND}. The high-velocity \sio\ tail is highlighted with white arrows.}
        \label{FIGURE:CHANNEL}
\end{figure*}

\textit{The Tail:} Extending $\sim0.044~\rm pc$ from the \hchii\ region center to the southwest, the tail approximately aligns with the arc symmetry axis and the most redshifted regions of \hchii\ region (Figs.~\ref{FIGURE:ATOMS-HII}a3 and \ref{FIGURE:QUARKS-HII-GAS}b). The PVDs along the tail axis for \ceo\ and \sio\ (Figs.~\ref{FIGURE:SiOC18OView}c1 and c2) reveal very high-velocity red line wings extending beyond $\rm v_{lsr}\approx 50$~\kms\ and $\rm v_{lsr}\approx 60$~\kms, respectively, at the position corresponding to the \hchii\ region ($\rm offset\sim0.75$\arcsec). Figure~\ref{FIGURE:CHANNEL} shows the channel map of \siofive, with a channel width of 2.4~\kms. The first-order moment maps of \ceo\ and \sio\ clearly show that the arc structure is blueshifted relative to the redshifted tail originating from the position of the \hchii\ region. The tail can still be seen up to $\rm v_{lsr}\sim60$~\kms\ in the channel map, indicating that the tail in the continuum is dominated by the high-velocity shocked molecular gas. 

\textit{The Cavity:} A cavity-like structure, characterized by 1.3~mm,  \ceo, and \sio\ emission, is outlined in Figs.~\ref{FIGURE:QUARKS-HII-EXTEND} and \ref{FIGURE:SiOC18OView} by the yellow dotted lines. Its edge is spatially associated with several faint low-mass cores with individual masses $\lesssim$ a few solar masses and total mass $\lesssim$ 24~\msun\ when assuming a \dustt\ of 20~K. These cores are identified using the source extraction algorithm \texttt{getsf} \citep{Menshchikov2021} by Jiao et al. in preparation. The two low-mass cores that account for only about 9\% of the arc flux density appear near the arc vertex, and therefore they are weak compared to the bright extended arc.

\subsection{Powering mechanism of the supersonic ionized gas}\label{subsec:mechanism}
In this section, we explore three possible origins of supersonic ionized gas by examining their compatibility with the observational signatures.

\textit{Trapped Collapsing Ionized Region:} 
Within an extremely dense \hchii\ region, a bounded structure and accretion of ionized gas could be achieved due to strong gravity, as proposed by \citet{Keto2003}. A large RRL optical depth accompanying the non-LTE effect would produce a net red-shifted profile for RRLs \citep{Peters2012,Klaassen2018}. 

However, this scenario is unlikely for our target. First, the \hchii\ region is too large for gravitational confinement. The maximum radius for a gravitationally trapped \hchii\ region, given by $r_{\rm max} = Gm_{*}/2c_{\rm i} \sim 60~\mathrm{au}$ \citep[where $m_{*}$ is mass of ionizing star;][]{Keto2003}, is significantly smaller than the observed size of 1560~au. Second, both \hthirtyalpha\ and \hfourtyalpha\ lines remain optically thin at $n_{\rm e} \sim 10^{5}~\rm cm^{-3}$ and the size scale of \hchii\ region. The optical depths of the RRL and the free-free continuum are given by:
\begin{align*}
&\tau_{\rm RRL} \sim 1.9\times10^3\,\left(\frac{T_{\rm e}}{{\rm K}}\right)^{-5/2}\,\left(\frac{EM}{{\rm cm^{-6}~pc}}\right)\,\left(\frac{\Delta \nu_{\rm RRL}}{{\rm kHz}}\right)^{-1} \\
&\tau_{\rm ff} \sim 3.3\times10^{-7}\,\left(\frac{T_{\rm e}}{{\rm 10^4~K}}\right)^{-1.35} \left(\frac{EM}{{\rm cm^{-6}~pc}}\right) \left(\frac{\nu_{\rm C}}{{\rm GHz}}\right)^{-2.1},  
\end{align*}
where $\nu_{\rm C}$ is the continuum frequency \citep{Wilson2013,Condon2016}. Both values are $<0.015$ for the targeted source. Non-LTE effects, such as pressure broadening, are negligible, as evidenced by the near-zero ratio of the Lorentzian line width to the thermal line width in frequency $\Delta \nu_{\rm L}/\Delta \nu_{\rm therm} = 1.2\,(n_{\rm e}/10^5~{\rm cm^{-3}})\,(N/92)^7 \sim0.007$, for $N = 40$ and $n_{\rm e}\sim2\times10^5$~\cmcube\ \citep{Keto2008}. 

Using \texttt{RADMC-3D} non-LTE radiative transfer code \citep{Dullemond2012}, we model an infalling isothermal \hchii\ region with a power-law density distribution $n_{\rm e} \propto r^{-3/2}$ derived from SED fitting in Sect.~\ref{subsec:hchii} (details in Appendix~\ref{APP:RADMC3D}). An infall velocity of ${\rm v}_{\rm e,0}^{\rm infall} = c_{\rm i}$ is adopted at its outer boundary, following \citet{Keto2003}, which assumes that the boundary corresponds to an R-critical jump where the ionized gas leaves the front at \ci. The modeled RRLs fail to reproduce the observed 20~\kms\ global velocity shift. We therefore conclude that a gravitationally bound, infalling ionized region is unlikely and that the shift reflects the actual kinematics of the ionized gas without significant bias from the optical effect.

\textit{Bow Shock of a Runaway High-Mass Star:} 
In this scenario, a high-mass star moves supersonically through its surrounding cloud. When the momentum flux in the stellar wind equals the ram pressure of the ambient medium, a bow-shaped ionized shock front is created ahead of the motion direction of the star and comoving with the star. Such runaway stars originate from either binary companion supernovae or cluster dynamical ejection \citep{Fujii2011, Rodriguez2020}. Although the arc detected in 1.3~mm continuum and molecular emission (\ceo\ and \sio) would initially suggest a bow shock from a runaway high-mass star and associated comoving ionized gas as the mechanism of supersonic ionized gas, we disfavor this interpretation for the following two reasons.

First, the observed velocity structure disagrees with the theoretical expectations. In the bow-shock scenario, the velocity of the ionized gas near the vertex should match stellar velocity $\rm v_*$, as the ionization front becomes trapped in the swept-up shell behind the bow shock formed ahead of the moving star \citep{Arthur2006}. In contrast, ionized gas in the trailing regions is expected to approach the ambient velocity due to deceleration and mixing with the surrounding medium \citep{vanBuren1992,Zhu2008}. Assuming that $\rm v_*$ equals the RRL velocity at the position closest to the arc vertex (${\rm v_{lsr}}\approx64$~\kms), the observed ionized-gas kinematics—appear redder toward the tail—seen in Figs.~\ref{FIGURE:ATOMS-HII}a3 and \ref{FIGURE:QUARKS-HII-GAS}b is inconsistent with this prediction, since the tail should exhibit velocities closer (bluer) to the ambient molecular gas. A more reliable evidence is the molecular gas kinematics because it is fully resolved. The bow-shock model predicts that molecular gas near the vertex should show velocities closest to $\rm v_*$, where the stellar wind imparts the greatest momentum along the direction of motion \citep{Wilkin1996,Zhu2015}. However, the parabolic velocity structure in the \ceo\ PVD (Fig.~\ref{FIGURE:SiOC18OView}b2) instead reliably reveals a blueshifted vertex relative to other parts of the arc—opposite to the model prediction.

Second, the stand-off distance is unlikely to be compatible with observations. The separation between the ionizing star and bow vertex derived from momentum conservation, known as the stand-off distance, is given by
\begin{equation*}
d_{\rm so} = \sqrt{\dot{m}_{\rm w}\,{\rm v_w}/4\pi\,\rho\,{\rm  v_{*}}^{2}},
\end{equation*}
where $\dot{m}_{\rm w}$ and $\rm v_{\rm w}$ are the stellar wind mass loss rate and terminal velocity, and $\rho$ is the ambient gas density \citep{Wilkin1996}. Adopting extreme parameters corresponding to an unusually strong stellar wind \citep[$\dot{m}_{\rm w} = 10^{-5}~\rm M_{\odot}~yr^{-1}$, $\rm v_{\rm w} = 3000$~\kms;][]{Kudritzki2000,Vink2001} and a lowest ambient density ($n_{\rm H} = 3\times10^{5}~\mathrm{cm^{-3}}$) corresponding to clump-scale density derived from ATLASGAL \citep{Urquhart2022}, the resulting maximum stand-off distance is $d_{\rm so} \sim 0.024~\mathrm{pc}$, close to the observed separation of $0.022~\mathrm{pc}$. However, in reality stellar winds are expected to be much weaker and the ambient density much higher, which reliably rules out a bow-shock origin for the arc. For example, a more reasonable parameter set of $\dot{m}_{\rm w} = 10^{-6}~\rm M_{\odot}~yr^{-1}$, $\rm v_{\rm w} = 2000$~\kms\ \citep{Zhang2024}, and $n_{\rm H} = 10^{6}~\mathrm{cm^{-3}}$ (typical of a dense core) yields only $d_{\rm so} \sim 0.003~\mathrm{pc}$.

\begin{figure}
\centering
\includegraphics[width=0.48\textwidth]{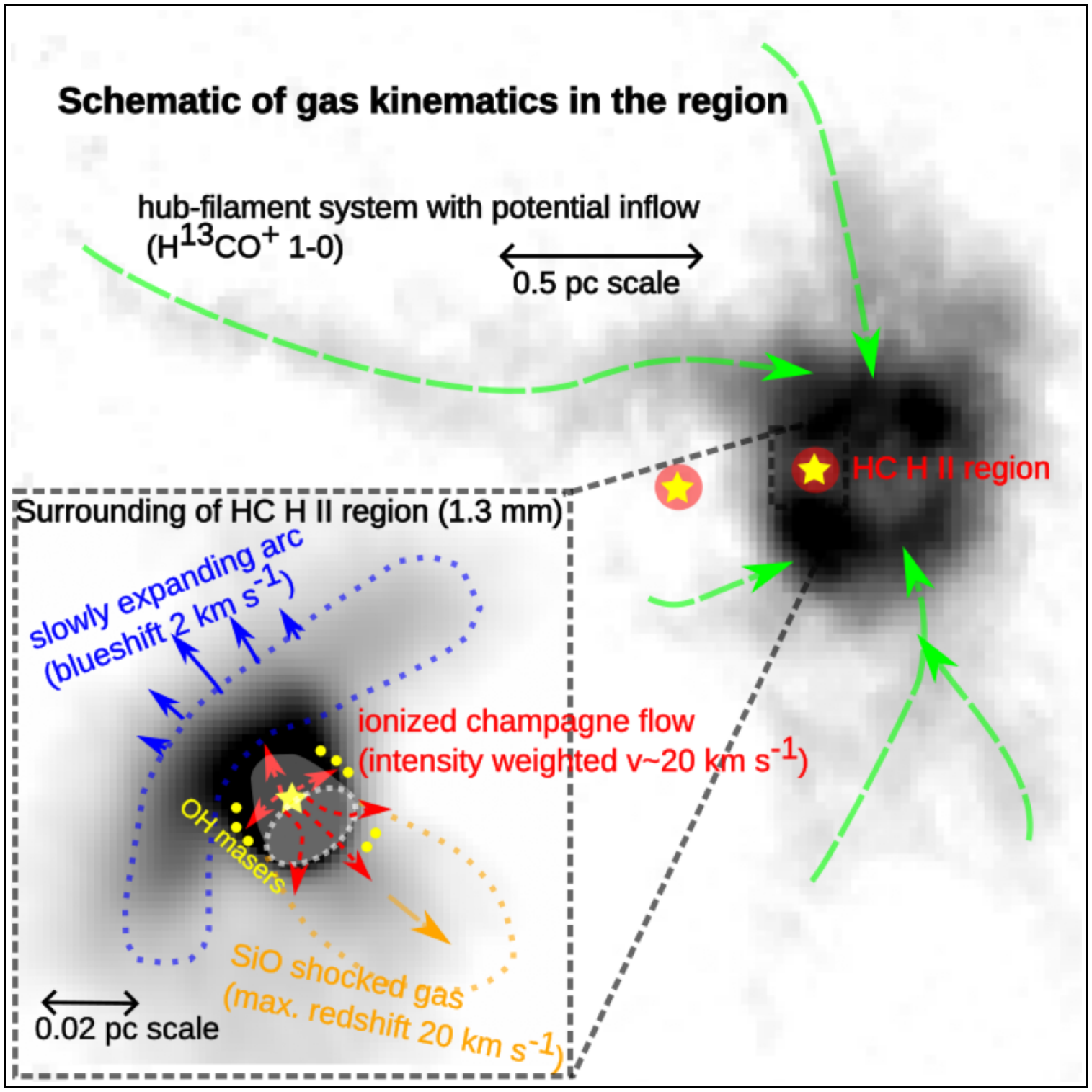} 
\caption{Schematic for the flows related with the \hchii\ region in this work.}
 \label{FIGURE:SCHEMATIC}
\end{figure}

\textit{Champagne Flow Breaking Out of a Hot Molecular Core:}
Having rejected both the bow shock and infalling ionized gas scenarios, we consider the champagne flow model where a stationary or slowly moving ionizing star drives supersonic ionized flows along density gradients, producing the observed global velocity shift. This model provides a better explanation for the observed kinematics. The arc structure is plausibly formed through compression of a pre-existing filament (likely embedding two low-mass cores) by the expanding \hchii\ region, where expansion toward the filament is severely inhibited by an upward density gradient. The expansion toward the upward density gradient tends to produce a parabolic density structure such as the exemplar shown in \citet{Zhu2015}. It is worth noting that two dense cores are likely located on either side of the arc’s vertex, suggesting that they may further constrain the direction of the expansion. The slow expansion ($\sim2$ to 3~\kms) produces the parabolic velocity structure with a blueshifted vertex ($\rm v_{\rm lsr} \sim 40$~\kms) in the \ceo\ PVD (Fig.~\ref{FIGURE:SiOC18OView}b2), and powers the narrow \sio\ shock emission (width $\sim$few \kms) observed around the clump velocity in the arc. In the opposite direction, a steep downward density gradient along the tail direction (from hot core to inter-core cavity shown in Fig.~\ref{FIGURE:QUARKS-HII-EXTEND}) enables ionized gas to flow outward at several times the sound speed $c_{\rm i}$, creating global redshifted emission \citep{Tenorio1979b}. A density contrast with a factor of a few hundred can create a maximum flow velocity of 3$c_{\rm i}$ to 4$c_{\rm i}$ \citep{Tenorio1979}. The outflowing ionized gas entrains or sweeps up the ambient molecular material, producing the redshifted high-velocity \sio\ shocked tail detected up to $\rm v_{\rm lsr} \sim 60$~\kms\ on the channel map (Fig.~\ref{FIGURE:CHANNEL}). Supporting evidence for the flow direction includes a possible velocity gradient in the RRL emission and the continuum/\sio\ tail observed along the NE–SW direction. This ``confined slow expansion in the head plus an accelerating champagne flow in the tail'' pattern, is reproduced in numerical simulations \citep{Bodenheimer1979,Tenorio1979b,Arthur2006,Zhu2015}, particularly when stellar winds assist expansion. Figure~\ref{FIGURE:SCHEMATIC} schematically illustrates this mass flow pattern for the I19095 \hchii\ region.

\begin{figure}
\centering
\includegraphics[width=0.48\textwidth]{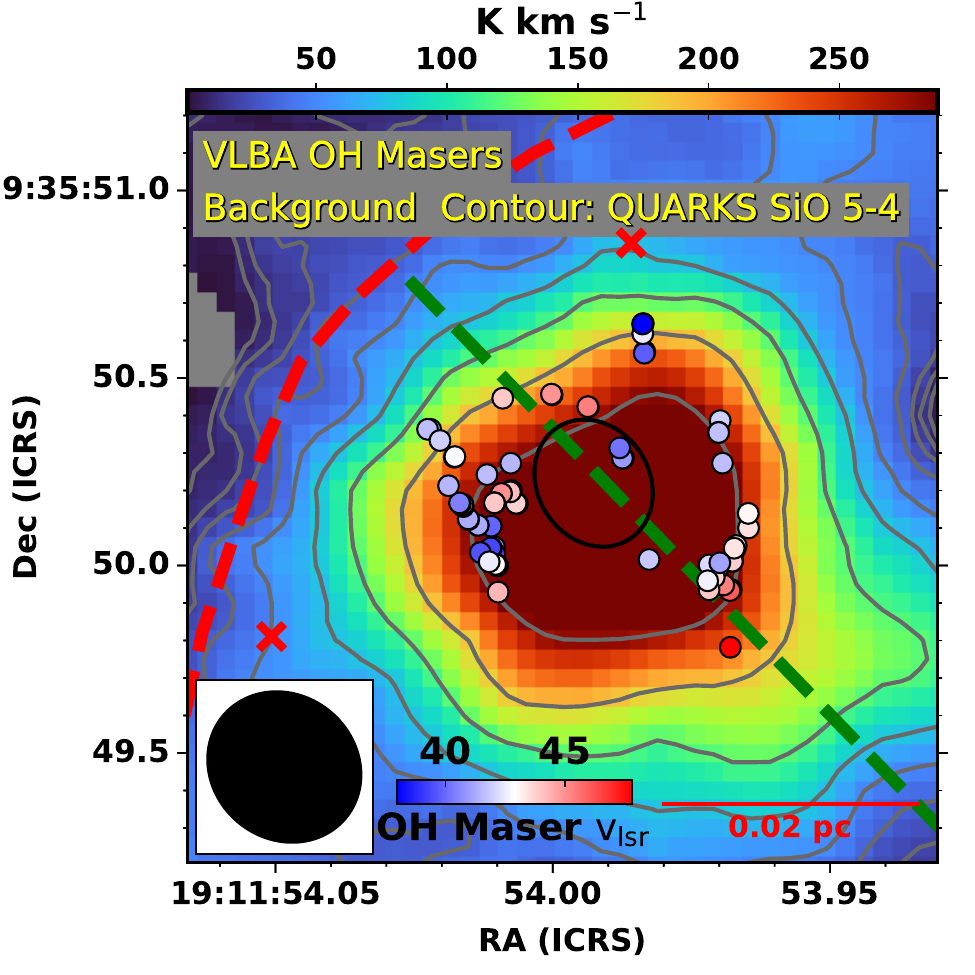} 
\caption{OH masers detected with the VLBA by \citet{Fish2005}, color-coded by their $\rm v_{\rm lsr}$. The background colormap represents a zoomed-in view of the QUARKS \siofive\ first-moment map shown in Fig.~\ref{FIGURE:SiOC18OView}a1.}
 \label{FIGURE:MASER}
\end{figure}

OH masers provide critical insight into the \hchii\ region--ambient gas interactions, tracing compressed shells between the fronts of ionization and shock \citep{Elitzur1978,Fish2007}. VLBA observations at 15~mas resolution show OH masers enveloping the \hchii\ region with ${\rm v}_{\mathrm{lsr}} = 39$ to 47~\kms, delineating a dense interface shell as shown in Fig.~\ref{FIGURE:MASER} \citep{Fish2005}. Critically, masers show no concentration around the arc, strongly disfavoring the bow shock model, which predicts maximal interaction in the bow structure and thus the highest probability of OH maser detection there \citep{Gasiprong2002}.

\begin{figure}
\centering
\includegraphics[width=0.48\textwidth]{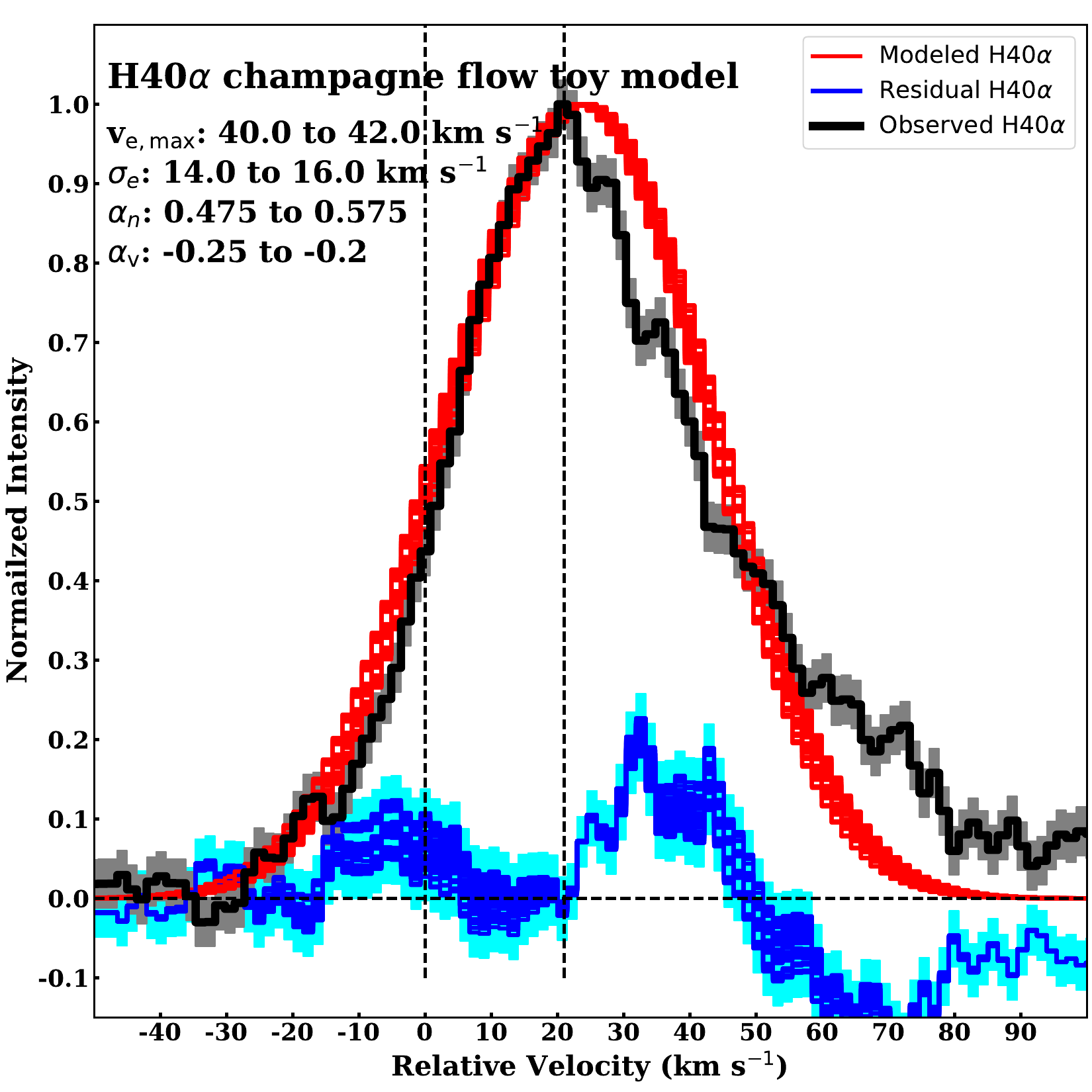} 
\caption{Comparison of the observed \hfourtyalpha\ line profile (black; gray shading indicates $1\sigma$ error) with the prediction of toy models of champagne flow (red). The residuals (model minus observation) are shown in blue, with cyan shading indicating the $1\sigma$ uncertainties. }
        \label{FIGURE:ToyModel}
\end{figure}

Given that QUARKS observations lack the resolution to probe the internal structure of the \hchii\ region whose size is partly evidenced by the OH maser distributions, constructing detailed morphological/kinematic models and performing comprehensive radiative transfer calculations \citep{Dullemond2012,Peters2012} for comparison with \hthirtyalpha\ observations would be overly speculative. Instead, we only test whether the global 20~\kms\ shift of the RRL can be reproduced by a simplified champagne flow toy model. Following \citet{Keto1995}, we define the ionized flow velocity, density, and cross-sectional area at position $z$ as ${\rm v_e}(z) = {\rm v_{e,0}^{flow}}(z/z_0)^{-\alpha_{\rm v}}$, $n_{\rm e}(z) = n_{{\rm e},0}^{\rm flow}(z/z_0)^{-\alpha_{n}}$, and $A(z) = A_0^{\rm flow}(z/z_0)^{-\alpha_A}$, respectively, with mass conservation requiring $\alpha_{\rm v} + \alpha_n + \alpha_A = 0$ ($\alpha_A < 0$, $\alpha_{\rm v} < 0$). In unresolved observations, the optically thin RRL profile is the integration of emission along the flow:
\begin{equation*}
I({\rm v}) \propto \int n_{\rm e}^2(z) A(z) \phi({\rm v_c} = {\rm v_e}(z), \sigma_{\rm e})\,dz,
\end{equation*}
where $\phi$ is the Gaussian profile centered at ${\rm v_e}(z)$ with dispersion $\sigma_{\rm e}$. Transforming to velocity space yields the following:
\begin{equation*}
I({\rm v}) \propto \int_0^{\rm v_{e,max}} {\rm v_e}^{(\alpha_{n} - 2\alpha_{\rm v} - 1)/\alpha_{\rm v}} \phi({\rm v_c} = {\rm v_e}, \sigma_{\rm e})\,d{\rm v_e}. 
\end{equation*}
We explore parameter grids: $\lvert\alpha_{\rm v}\rvert,\,\alpha_n \in [0.2,\,3.0]$ with 0.025 step, $\sigma_{\rm e} \in [12,\,25]$~\kms\ and $\rm v_{\mathrm{e},\max} \in [40,\,80]$~\kms\ with 1~\kms\ step. Models with a coefficient of determination $R^2 > 99.999$th percentile across all model grids shown in Fig.~\ref{FIGURE:ToyModel} reproduce well the 20~\kms\ global red shift and blue parts of the line profile. This toy model demonstrates the \textit{feasibility} of champagne flow origins for the global velocity shift, although it is not a fully complete model that can be comprehensively compared with observations. Deviations occur in the red wing, possibly due to over-simplified kinematics, mass distribution, and morphology. For instance, excess emission at relative velocity $>60$~\kms\ is likely associated with unmodeled high-velocity components beyond $\rm v_{e,max}$, while the deficit at 20--40~\kms\ likely indicates a central wind-cleared cavity with low density and velocity.

\section{Hatching out of the egg phase and Conclusions} \label{sec:discussion}
We evaluated three scenarios for the global supersonic velocity shift in the optically thin RRLs of I19095 \hchii\ region and concluded that a champagne flow represents the most plausible mechanism. The \hchii\ region remains deeply embedded within a 0.02 pc-scale hot molecular core, indicating that we are observing a rapid \textit{``hatching out of the egg''} transition phase in the evolution of the hot core--\hchii\ region: the ionized gas has just broken out of the parental hot core and flows supersonically into the lower-density inter-core region of the clump. The potential inclination of the flow implies a global velocity $>20$~\kms. The slightly more evolved nature of the low-density \hchii\ regions proposed by \citet{RiveraSoto2020} is consistent with the interpretation that the I19095 \hchii\ region is undergoing the ``hatching'' phase, resulting in its lower electron density compared to the more embedded, hyperdense \hchii\ regions.

In subsequent evolution, the ionization front is expected to expand supersonically into the inter-core regions within the clump and form an irregular \uchii\ region, accompanied with flow deceleration, which will diminish the global velocity shift and thus make the detection window of ``hatching'' phase extremely brief. The rarity of supersonic velocity shifts among \hchii\ regions (1/16 in the ATOMS survey; red dots in Fig.~\ref{fig:velo-comparison}c) reflects that the short duration of this phase likely constitutes $<10\%$ of the typical \hchii\ region lifetime. Since these supersonic flows precede the formation of \uchii\ regions, the “hatching” phase must be much shorter than the typical \uchii\ region lifetime of $\sim3\times10^5$~yr \citep{Wood1989, Churchwell2002, Mottram2011, Kalcheva2018}. Therefore, the maximum timescale of the “hatching” phase, estimated by combining the \uchii\ region lifetime with the number ratio of supersonic \hchii\ regions to all \hchii\ regions in the ATOMS sample (1/16), is approximately $1/16\times3\times10^5\approx2\times10^4$~yr. Meanwhile, using the ratio of supersonic \hchii\ regions to all compact (UC + \hchii) regions with radii $<0.05$~pc (1/63), the typical timescale of the “hatching” phase is about $1/63\times3\times10^5\approx5\times10^3$~yr. \textit{Given that the Galactic OB star formation rate is around 0.01 to 0.05 per year, only tens of \hchii\ regions in such ``hatching'' phase probably exist in the Galaxy today, highlighting the extreme difficulties in detection of such exotic \hchii\ regions} \citep{Reed2005, Robitaille2010, Chomiuk2011}.

Although extremely short compared to the HMSF timescales, this phase—and its associated supersonic champagne flow—should prevail during the transitions from \hchii\ regions embedded in hot cores to \uchii\ regions \citep{Kim2001}. This prevalence stems from the inherent anisotropic density distributions within the parental hot cores and their surroundings, as demonstrated in the DIHCA project \citep[Digging into the Interior of Hot Cores with ALMA;][]{Olguin2021,Olguin2022,Olguin2023}. Such anisotropy naturally drives gradient-driven champagne flows. Asymmetric expansion beginning at the \hchii\ region stage complicates the expansion process and contributes to the lifetime-dynamical time discrepancy in \uchii\ region \citep{Wood1989, Kim2001}.

Stellar winds significantly influence early \hii\ region kinematics \citep{Geen2021}. Simulations show that winds combined with density gradient-driven champagne flows produce cavities and complex velocity structures \citep{Arthur2006}. Our $\sim2000$~au resolution QUARKS data cannot resolve the inner structure of I19095 \hchii\ region, and discrepancies between continuum SED fitting and RRL modeling power-law indices likely arise from differing morphological assumptions and wind-induced complexity.

We advocate for future searches for the supersonic \hchii\ region coupled with long baseline ALMA observations resolving to $\sim100$~au scales, to map the gas dynamics during this critical and rapid transient phase.

\begin{acknowledgments}
SZ gratefully acknowledges support by the CAS-ANID project CAS220003. GG and LB gratefully acknowledge support by the ANID BASAL project FB210003. AS gratefully acknowledges support by the Fondecyt Regular (project code 1220610), and ANID BASAL project FB210003. This research was carried out in part at the Jet Propulsion Laboratory, which is operated by the California Institute of Technology under a contract with the National Aeronautics and Space Administration (80NM0018D0004). We want to thank the anonymous referee for the constructive comments that helped to improve the quality of the paper.

This paper makes use of the following ALMA data: ADS/JAO.ALMA\#2019.1.00685.S, 2021.1.00095.S and 2023.1.00425. ALMA is a partnership of ESO (representing its member states), NSF (USA), and NINS (Japan), together with NRC (Canada), MOST and ASIAA (Taiwan), and KASI (Republic of Korea), in cooperation with the Republic of Chile. The Joint ALMA Observatory is operated by ESO, AUI/NRAO, and NAOJ.
\end{acknowledgments}

%

\vspace{5mm}
\facilities{ALMA.}


\software{\texttt{astropy} \citep{Astropy2022}, \texttt{CASA} \citep{CASA2022}, \texttt{FilFinder} \citep{Koch2015}, \texttt{XCLASS} \citep{XCLASS2017}, \texttt{RADMC-3D} \citep{Dullemond2012}. }



\appendix
\section{\texorpdfstring{\hthirtyalpha}{} line deblending}\label{APP:XCLASS}
The philosophy of \hthirtyalpha\ line deblending is to remove the overlapping molecular lines by fitting their multiple transitions outside the \hthirtyalpha\ frequency range. Our general workflow includes three steps: line identification, \texttt{XCLASS} model fitting, and individual line reproduction.

First, we select the northern arc where complex organic molecules are abundant while \hthirtyalpha\ is absent. Based on the CDMS database\footnote{\href{https://cdms.astro.uni-koeln.de/}{https://cdms.astro.uni-koeln.de/}}, we have identified molecular species with more than one transitions: $^{13}$CH$_3$CN, CH$_3$OH, CH$_3^{18}$OH;A/E, CH$_3$CHO;v=0, CH$_3$OCHO;v18=1, C$_2$H$_5$OH, CH$_3$C$^{15}$N, C$_2$H$_5$CN, CH$_3$OCH$_3$, SO$_2$;v2=1, and HC(O)NH$_2$. In addition, there are five species with only one transition: $^{13}$CS, OCS, t-HCOOH, and HNCO. By fitting observations with the \texttt{myXCLASSFit} algorithm, we obtain a template model for further fitting.

\begin{figure*}
\centering
\includegraphics[width=\linewidth]{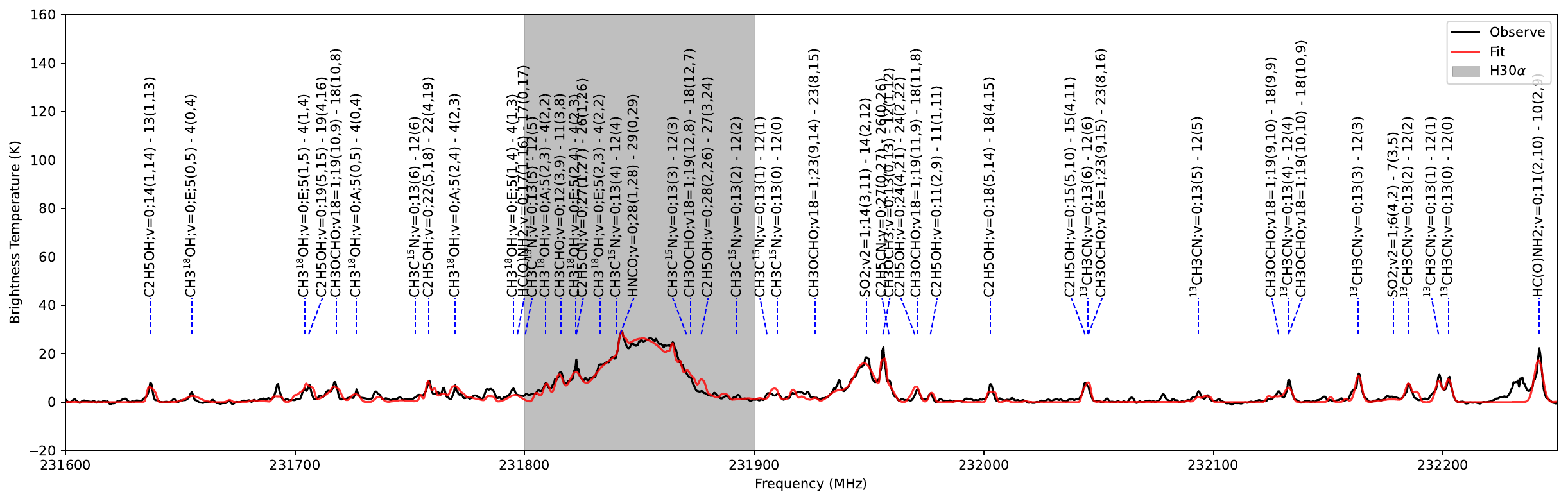}
\caption{Example of successfully fitted spectrum towards the 1.3~mm peak where the \hchii\ region is located. Black and red lines are the observed and fitted spectra. Gray shade highlights the frequency range of \hthirtyalpha.}
\label{FIGURE:LABEL}
\end{figure*}

Then, we add \hthirtyalpha\ to the model, which can be described by one kinematic component characterized by source size, \te, electron EM, Gaussian $\Delta \rm v_G$ and Lorentzian $\Delta\rm v_L$ line width contributions, and velocity offset $\rm v_{\rm off}$. In the region with strong \hthirtyalpha\ detection, we then apply the model to fit the spectra pixel by pixel with \texttt{myXCLASSMapFit}. As an example, we show a successfully fitted spectrum in Fig.~\ref{FIGURE:LABEL}.

Finally, we retrieve the individual model \hthirtyalpha\ to reproduce the RRLs in each pixel. The fitted EM, centroid velocity and line width are shown in Fig.~\ref{FIGURE:QUARKS-HII-GAS}(a-c).

\section{\hfourtyalpha\ line profile} \label{APP:H40aProfile}
\begin{figure*}
\centering
\includegraphics[width=0.75\linewidth]{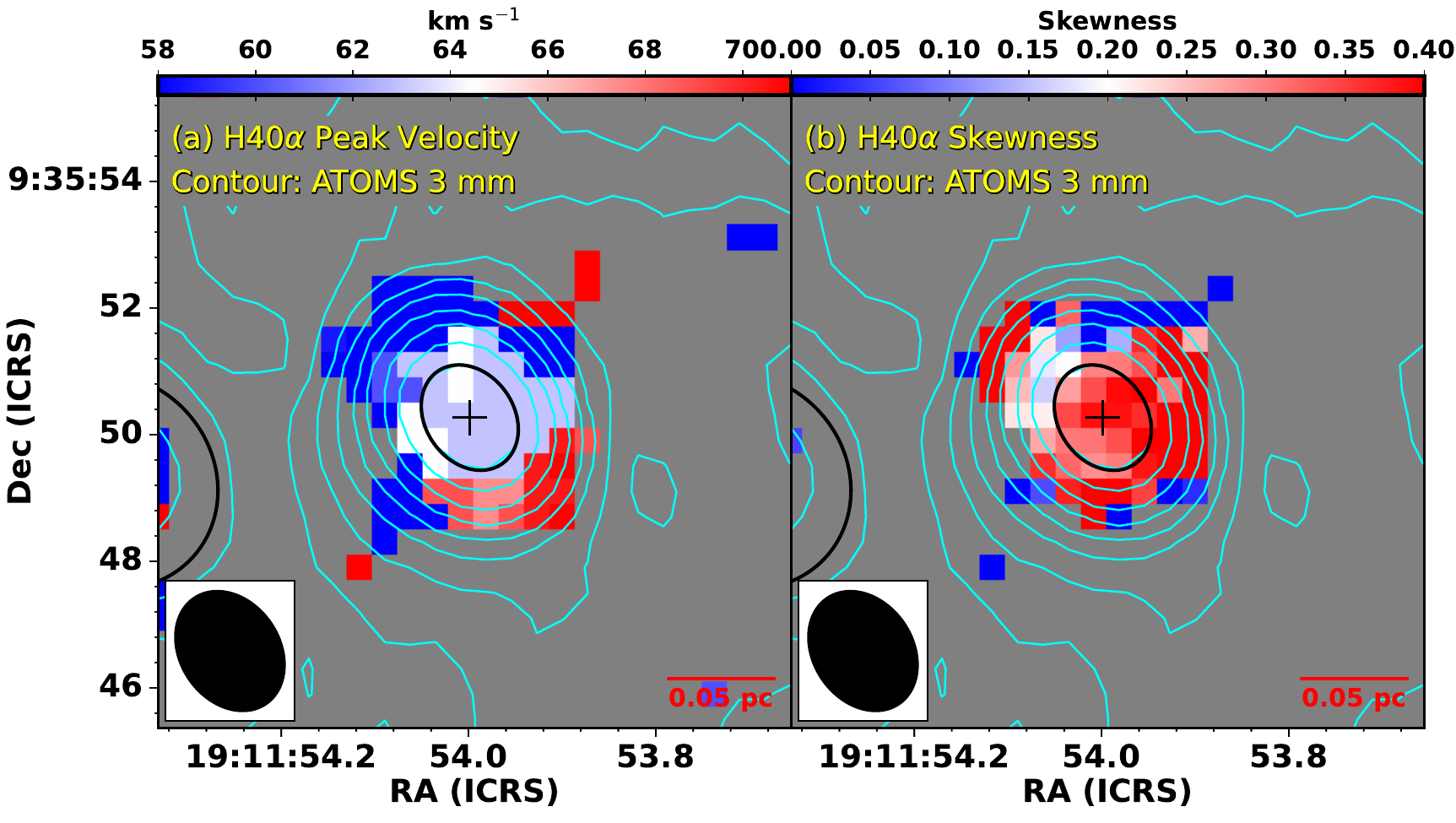}
\caption{Peak velocity and skewness maps for \hfourtyalpha, overlaid with ATOMS 3~mm emission contours with levels the same as Fig.~\ref{FIGURE:ATOMS-HII}.}
\label{FIGURE:H40aProfile}
\end{figure*}

Figure~\ref{FIGURE:H40aProfile} shows the maps of the peak velocity and skewness for \hfourtyalpha. Global changes along the NE-SW direction are obvious.

\section{SED fitting}\label{APP:SED}
The SED fitting methodology follows \citet{Keto2003}, modeling free-free emission from a spherical \hchii\ region as:
\begin{equation}
S_{\rm \nu} = \frac{4\pi k_{\rm B} T_{\rm e} \nu^{2}}{c^{2}} \int_{0}^{\theta_{0}} \theta \left(1 - e^{-\tau_{\rm ff}}\right)  d\theta,
\end{equation}
where $\theta = \sqrt{x^{2} + y^{2}}/D$ is the angular sky coordinate and $\tau_{\rm ff}(T_{\rm e}, EM, \nu)$ denotes the frequency-dependent free-free optical depth (Sect.~\ref{subsec:mechanism}). Varying electron density profiles produce distinct SED shapes. 
We adopt $T_{\rm e} = 10^{4}$~K and fit both uniform-density and power-law ($n_{\rm e} \propto r^{-3/2}$) models, employing the simplifications detailed in \citet{Keto2003}. The fitting incorporates archival VLA data spanning 7~mm to 6~cm:

\textit{7~mm:} Observations of I19095 were conducted in the VLA B-configuration at $Q$-band (Project ID: 23A-205; PI: Fengwei Xu). The WIDAR correlator provided 8~GHz continuum coverage centered at 44~GHz. After manual calibration in \texttt{CASA} 6.5.4, four sub-band images were generated at 41.0, 43.1, 45.4, and 47.5~GHz, enabled by the upgraded correlator. Self-calibration improved the image quality, yielding a synthesized beam of $\sim0.25$\arcsec$\times0.18$\arcsec\ and an image rms noise of $\sim0.4$~\mjybeam. The measured flux densities are $170.7 \pm 2.1$, $167.3 \pm 1.5$, $167.6 \pm 1.8$, and $164.0 \pm 1.8$~mJy, respectively, with a source size of $\sim0.2$\arcsec$\times$0.16\arcsec. 

\textit{1.3~cm \& 6.0~cm:} Data from \citet{Kavak2021} (B-configuration; $K$- and $C$-bands) yield flux densities of $130.87 \pm 2.60$~mJy and $39.50 \pm 1.60$~mJy, respectively. The synthesized beam sizes are 0.91\arcsec$\times$0.84\arcsec\ and 1.49\arcsec$\times$1.28\arcsec, with corresponding source sizes of 0.34\arcsec$\times$0.23\arcsec\ and 1.18\arcsec$\times$0.26\arcsec, respectively. 

\textit{2.0~cm \& 3.6~cm:} Archival B-configuration data from \citet{Kurtz1994} provide flux densities of $107.8 \pm 1.8$~mJy and $62.6 \pm 0.6$~mJy. 
The synthesized beam sizes are 0.54\arcsec$\times$0.46\arcsec\ and 1.02\arcsec$\times$0.77\arcsec, with source sizes of 0.6\arcsec$\times$0.5\arcsec\ and 1.1\arcsec$\times$0.9\arcsec, respectively.

\section{Position-velocity diagrams along the arc} \label{APP:PVDs}
\begin{figure*}[htb!]
\centering
\includegraphics[width=0.97\textwidth]{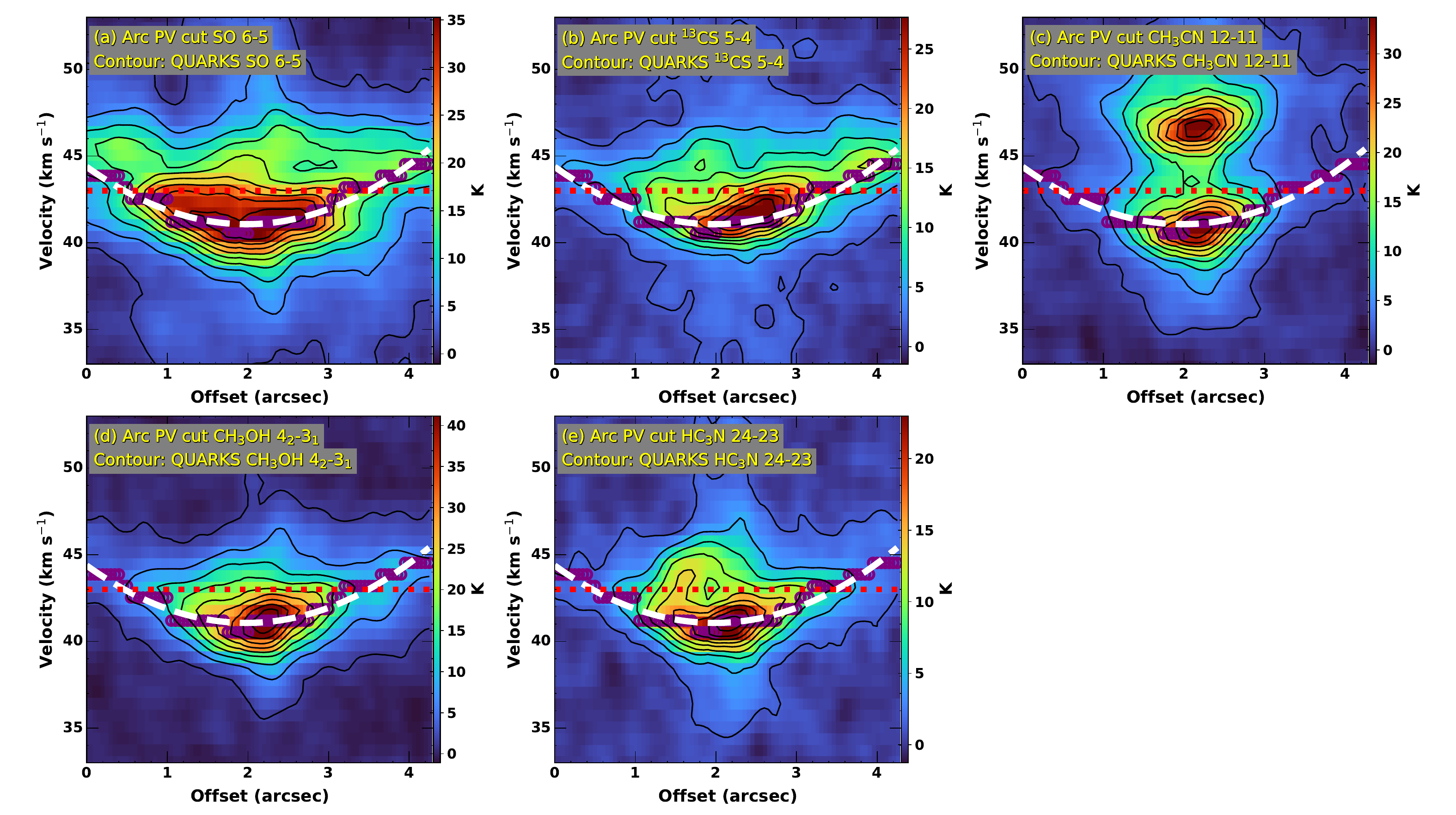}
       \caption{Position-velocity diagrams along the arc. Panels (a)-(e) show the PVDs of SO $6-5$, \tcs\ $5-4$, \chtcn\ $12-11$, \chtoh\ $4_2-3_1$, and HC$_3$N $24-23$, respectively. Contours run from 3$\sigma$ to 37.2, 30.6, 36.3, 47.3, and 28.2 K in 10 linear steps, respectively. Other labels and lines are as in Fig.~\ref{FIGURE:SiOC18OView}.}
        \label{FIGURE:ALL_PV}
\end{figure*}

Figure~\ref{FIGURE:ALL_PV} shows additional PVDs of several QUARKS molecular lines. The \ceo-fitted parabola matches the velocity structures well, strongly supporting the existence of parabolic kinematics.

\section{\texorpdfstring{\texttt{RADMC-3D}}{} modeling of an infalling \texorpdfstring{\hchii}{} region} \label{APP:RADMC3D}
\begin{figure*}[htb!]
\centering
\includegraphics[width=0.95\textwidth]{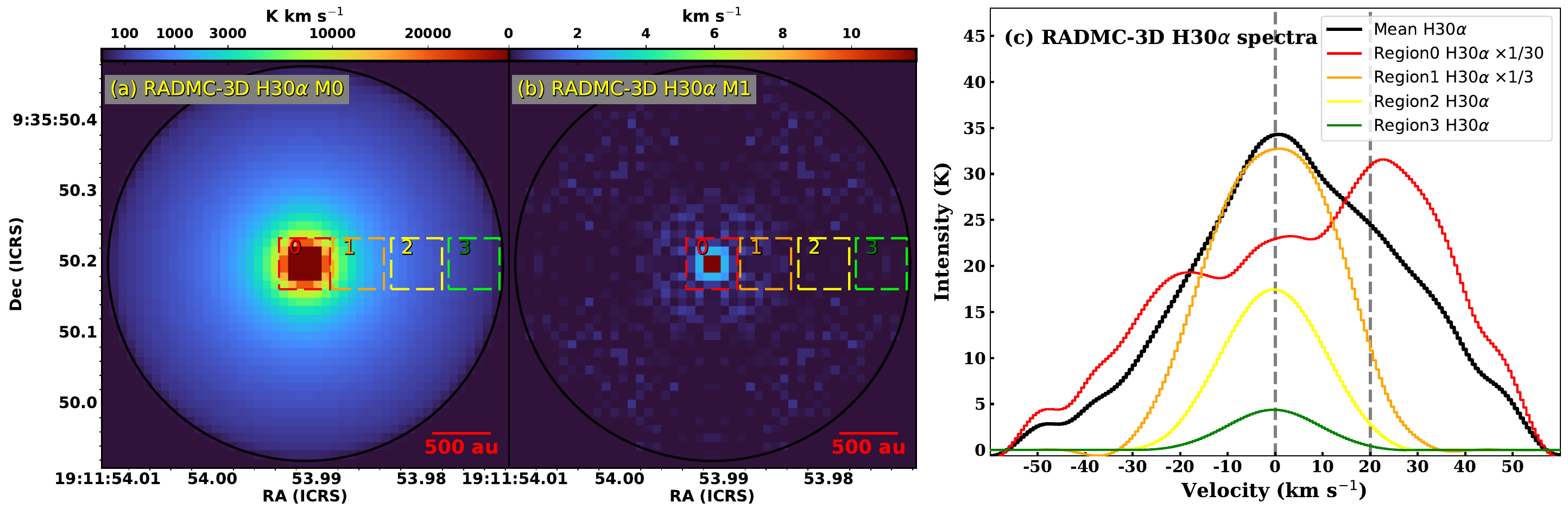}
       \caption{\hthirtyalpha\ \texttt{RADMC-3D} model of an infalling \hchii\ region. (a and b) 0th and 1st moment maps. (c) averaged spectra over the entire region and four subregions. For clarity in the figure, the line profiles of Regions 0 and 1 have been divided by factors of 30 and 3, respectively.}
        \label{FIGURE:RADMC3D}
\end{figure*}

With the non-LTE radiative transfer code \texttt{RADMC-3D} specifically modified for RRLs by \citet{Peters2012}, we modeled \hthirtyalpha\ lines of an \hchii\ region using power-law density and velocity distributions consistent with mass conservation: $n_{\rm e} = {n}_{\rm e,0}^{\rm infall} \left(r/r_{\rm 0}\right)^{-3/2}~ \left(r_{i}<r<r_{0}\right)$ and ${\rm v}_{\rm e} = -{\rm v}_{\rm e,0}^{\rm infall} \left(r/r_{\rm 0}\right)^{-1/2}~ \left(r_{i}<r<r_{0}\right)$. The outer radius $r_{\rm 0} = 1650$~au and the density 
$n_{\rm e,0}^{\rm infall} = 0.7\times10^5$~\cmcube\ were adopted from the free-free SED fitting results. We set the radius of the density plateau to $r_{i} = r_{\rm 0}/50 \sim33$~au. The infall velocity at the outer boundary of \hchii\ region, ${\rm v}_{\rm e,0}^{\rm infall}$, was assumed to be the sound speed $c_{\rm i}$. The model utilized a 3D grid of $150\times150\times150$ cells with 
10$^6$ photons packages and a constant $T_{\rm e} = 10^4$~K. Figure~\ref{FIGURE:RADMC3D} presents the resulting 0th and 1st moment maps, along with the spatially averaged spectra over the entire \hchii\ region and four subregions. While the source-averaged \hthirtyalpha\ line exhibits an asymmetric profile, primarily due to emission from the innermost pixels (Region 0), the model \hthirtyalpha\ lines show no significant redshift, except for the innermost pixels. The small bumps seen in the line profile of Region 0 arise because the line shape varies rapidly from pixel to pixel in this region; thus, when only a limited number of pixels are averaged, the resulting spectrum is not fully smooth.

\bibliography{sample631}{}

@ARTICLE{ATOMSI,
       author = {{Liu}, Tie and {Evans}, Neal J. and {Kim}, Kee-Tae and {Goldsmith}, Paul F. and {Liu}, Sheng-Yuan and {Zhang}, Qizhou and {Tatematsu}, Ken'ichi and {Wang}, Ke and {Juvela}, Mika and {Bronfman}, Leonardo and {Cunningham}, Maria R. and {Garay}, Guido and {Hirota}, Tomoya and {Lee}, Jeong-Eun and {Kang}, Sung-Ju and {Li}, Di and {Li}, Pak-Shing and {Mardones}, Diego and {Qin}, Sheng-Li and {Ristorcelli}, Isabelle and {Tej}, Anandmayee and {Toth}, L. Viktor and {Wu}, Jing-Wen and {Wu}, Yue-Fang and {Yi}, Hee-weon and {Yun}, Hyeong-Sik and {Liu}, Hong-Li and {Peng}, Ya-Ping and {Li}, Juan and {Li}, Shang-Huo and {Lee}, Chang Won and {Shen}, Zhi-Qiang and {Baug}, Tapas and {Wang}, Jun-Zhi and {Zhang}, Yong and {Issac}, Namitha and {Zhu}, Feng-Yao and {Luo}, Qiu-Yi and {Soam}, Archana and {Liu}, Xun-Chuan and {Xu}, Feng-Wei and {Wang}, Yu and {Zhang}, Chao and {Ren}, Zhiyuan and {Zhang}, Chao},
        title = "{ATOMS: ALMA Three-millimeter Observations of Massive Star-forming regions - I. Survey description and a first look at G9.62+0.19}",
      journal = {\mnras},
         year = 2020,
        month = aug,
       volume = {496},
       number = {3},
        pages = {2790-2820},
          doi = {10.1093/mnras/staa1577},
archivePrefix = {arXiv},
       eprint = {2006.01549},
 primaryClass = {astro-ph.GA},
       adsurl = {https://ui.adsabs.harvard.edu/abs/2020MNRAS.496.2790L}
}

@ARTICLE{ATOMSIII,
       author = {{Liu}, Hong-Li and {Liu}, Tie and {Evans}, Neal J., II and {Wang}, Ke and {Garay}, Guido and {Qin}, Sheng-Li and {Li}, Shanghuo and {Stutz}, Amelia and {Goldsmith}, Paul F. and {Liu}, Sheng-Yuan and {Tej}, Anandmayee and {Zhang}, Qizhou and {Juvela}, Mika and {Li}, Di and {Wang}, Jun-Zhi and {Bronfman}, Leonardo and {Ren}, Zhiyuan and {Wu}, Yue-Fang and {Kim}, Kee-Tae and {Lee}, Chang Won and {Tatematsu}, Ken'ichi and {Cunningham}, Maria R. and {Liu}, Xun-Chuan and {Wu}, Jing-Wen and {Hirota}, Tomoya and {Lee}, Jeong-Eun and {Li}, Pak-Shing and {Kang}, Sung-Ju and {Mardones}, Diego and {Ristorcelli}, Isabelle and {Zhang}, Yong and {Luo}, Qiu-Yi and {Toth}, L. Viktor and {Yi}, Hee-weon and {Yun}, Hyeong-Sik and {Peng}, Ya-Ping and {Li}, Juan and {Zhu}, Feng-Yao and {Shen}, Zhi-Qiang and {Baug}, Tapas and {Dewangan}, L.~K. and {Chakali}, Eswaraiah and {Liu}, Rong and {Xu}, Feng-Wei and {Wang}, Yu and {Zhang}, Chao and {Li}, Jinzeng and {Zhang}, Chao and {Zhou}, Jianwen and {Tang}, Mengyao and {Xue}, Qiaowei and {Issac}, Namitha and {Soam}, Archana and {{\'A}lvarez-Guti{\'e}rrez}, Rodrigo H.},
        title = "{ATOMS: ALMA three-millimeter observations of massive star-forming regions - III. Catalogues of candidate hot molecular cores and hyper/ultra compact H II regions}",
      journal = {\mnras},
         year = 2021,
        month = aug,
       volume = {505},
       number = {2},
        pages = {2801-2818},
          doi = {10.1093/mnras/stab1352},
archivePrefix = {arXiv},
       eprint = {2105.03554},
 primaryClass = {astro-ph.GA},
       adsurl = {https://ui.adsabs.harvard.edu/abs/2021MNRAS.505.2801L}
}

@ARTICLE{ATOMSVIII,
       author = {{Qin}, Sheng-Li and {Liu}, Tie and {Liu}, Xunchuan and {Goldsmith}, Paul F. and {Li}, Di and {Zhang}, Qizhou and {Liu}, Hong-Li and {Wu}, Yuefang and {Bronfman}, Leonardo and {Juvela}, Mika and {Lee}, Chang Won and {Garay}, Guido and {Zhang}, Yong and {He}, Jinhua and {Hsu}, Shih-Ying and {Shen}, Zhi-Qiang and {Lee}, Jeong-Eun and {Wang}, Ke and {Tang}, Ningyu and {Tang}, Mengyao and {Zhang}, Chao and {Yue}, Yinghua and {Xue}, Qiaowei and {Li}, Shanghuo and {Peng}, Yaping and {Dutta}, Somnath and {Ge}, Jixing and {Xu}, Fengwei and {Chen}, Long-Fei and {Baug}, Tapas and {Dewangan}, Lokesh and {Tej}, Anandmayee},
        title = "{ATOMS: ALMA Three-millimeter Observations of Massive Star-forming regions - VIII. A search for hot cores by using C$_{2}$H$_{5}$CN, CH$_{3}$OCHO, and CH$_{3}$OH lines}",
      journal = {\mnras},
         year = 2022,
        month = apr,
       volume = {511},
       number = {3},
        pages = {3463-3476},
          doi = {10.1093/mnras/stac219},
archivePrefix = {arXiv},
       eprint = {2201.10044},
 primaryClass = {astro-ph.GA},
       adsurl = {https://ui.adsabs.harvard.edu/abs/2022MNRAS.511.3463Q}
}

@ARTICLE{QUARKSI,
       author = {{Liu}, Xunchuan and {Liu}, Tie and {Zhu}, Lei and {Garay}, Guido and {Liu}, Hong-Li and {Goldsmith}, Paul and {Evans}, Neal and {Kim}, Kee-Tae and {Liu}, Sheng-Yuan and {Xu}, Fengwei and {Lu}, Xing and {Tej}, Anandmayee and {Mai}, Xiaofeng and {Bronfman}, Leonardo and {Li}, Shanghuo and {Mardones}, Diego and {Stutz}, Amelia and {Tatematsu}, Ken'ichi and {Wang}, Ke and {Zhang}, Qizhou and {Qin}, Sheng-Li and {Zhou}, Jianwen and {Luo}, Qiuyi and {Zhang}, Siju and {Cheng}, Yu and {He}, Jinhua and {Gu}, Qilao and {Li}, Ziyang and {Zhang}, Zhenying and {Zhang}, Suinan and {Saha}, Anindya and {Dewangan}, Lokesh and {Sanhueza}, Patricio and {Shen}, Zhiqiang},
        title = "{The ALMA-QUARKS Survey. I. Survey Description and Data Reduction}",
      journal = {Research in Astronomy and Astrophysics},
         year = 2024,
        month = feb,
       volume = {24},
       number = {2},
          eid = {025009},
        pages = {025009},
          doi = {10.1088/1674-4527/ad0d5c},
archivePrefix = {arXiv},
       eprint = {2311.08651},
 primaryClass = {astro-ph.GA},
       adsurl = {https://ui.adsabs.harvard.edu/abs/2024RAA....24b5009L}
}

@ARTICLE{QUARKSII,
       author = {{Xu}, Fengwei and {Wang}, Ke and {Liu}, Tie and {Zhu}, Lei and {Garay}, Guido and {Liu}, Xunchuan and {Goldsmith}, Paul and {Zhang}, Qizhou and {Sanhueza}, Patricio and {Qin}, Shengli and {He}, Jinhua and {Juvela}, Mika and {Tej}, Anandmayee and {Liu}, Hongli and {Li}, Shanghuo and {Morii}, Kaho and {Zhang}, Siju and {Zhou}, Jianwen and {Stutz}, Amelia and {Evans}, Neal J. and {Kim}, Kee-Tae and {Liu}, Shengyuan and {Mardones}, Diego and {Li}, Guangxing and {Bronfman}, Leonardo and {Tatematsu}, Ken'ichi and {Lee}, Chang Won and {Lu}, Xing and {Mai}, Xiaofeng and {Jiao}, Sihan and {Chibueze}, James O. and {Su}, Keyun and {T{\'o}th}, Viktor L.},
        title = "{The ALMA-QUARKS Survey. II. The ACA 1.3 mm Continuum Source Catalog and the Assembly of Dense Gas in Massive Star-Forming Clumps}",
      journal = {Research in Astronomy and Astrophysics},
         year = 2024,
        month = jun,
       volume = {24},
       number = {6},
          eid = {065011},
        pages = {065011},
          doi = {10.1088/1674-4527/ad3dc3},
archivePrefix = {arXiv},
       eprint = {2404.02275},
 primaryClass = {astro-ph.GA},
       adsurl = {https://ui.adsabs.harvard.edu/abs/2024RAA....24f5011X}
}

@INPROCEEDINGS{Kurtz2005,
       author = {{Kurtz}, Stan},
        title = "{Hypercompact HII regions}",
    booktitle = {Massive Star Birth: A Crossroads of Astrophysics},
         year = 2005,
       editor = {{Cesaroni}, R. and {Felli}, M. and {Churchwell}, E. and {Walmsley}, M.},
       volume = {227},
        month = jan,
        pages = {111-119},
       series = {},
          doi = {10.1017/S1743921305004424},
       adsurl = {https://ui.adsabs.harvard.edu/abs/2005IAUS..227..111K}
}

@ARTICLE{Moscadelli2021,
       author = {{Moscadelli}, L. and {Cesaroni}, R. and {Beltr{\'a}n}, M.~T. and {Rivilla}, V.~M.},
        title = "{The ionized heart of a molecular disk. ALMA observations of the hyper-compact HII region G24.78+0.08 A1}",
      journal = {\aap},
         year = 2021,
        month = jun,
       volume = {650},
          eid = {A142},
        pages = {A142},
          doi = {10.1051/0004-6361/202140829},
archivePrefix = {arXiv},
       eprint = {2105.01516},
 primaryClass = {astro-ph.GA},
       adsurl = {https://ui.adsabs.harvard.edu/abs/2021A&A...650A.142M}
}

@ARTICLE{Miyawaki2023,
       author = {{Miyawaki}, Ryosuke and {Hayashi}, Masahiko and {Hasegawa}, Tetsuo},
        title = "{An expanding ring of the hypercompact H II region W 49 N:A2}",
      journal = {\pasj},
         year = 2023,
        month = feb,
       volume = {75},
       number = {1},
        pages = {225-232},
          doi = {10.1093/pasj/psac105},
archivePrefix = {arXiv},
       eprint = {2212.01999},
 primaryClass = {astro-ph.SR},
       adsurl = {https://ui.adsabs.harvard.edu/abs/2023PASJ...75..225M}
}

@ARTICLE{Guzman2020,
       author = {{Guzm{\'a}n}, Andr{\'e}s E. and {Sanhueza}, Patricio and {Zapata}, Luis and {Garay}, Guido and {Rodr{\'\i}guez}, Luis Felipe},
        title = "{A Photoionized Accretion Disk around a Young High-mass Star}",
      journal = {\apj},
         year = 2020,
        month = nov,
       volume = {904},
       number = {1},
          eid = {77},
        pages = {77},
          doi = {10.3847/1538-4357/abbe09},
archivePrefix = {arXiv},
       eprint = {2010.00244},
 primaryClass = {astro-ph.SR},
       adsurl = {https://ui.adsabs.harvard.edu/abs/2020ApJ...904...77G}
}

@ARTICLE{Komesh2024,
       author = {{Komesh}, Toktarkhan and {Garay}, Guido and {Henkel}, Christian and {Omar}, Aruzhan and {Estalella}, Robert and {Assembay}, Zhandos and {Li}, Dalei and {Guzm{\'a}n}, Andr{\'e}s and {Esimbek}, Jarken and {Huang}, Jiasheng and {He}, Yuxin and {Alimgazinova}, Nazgul and {Kyzgarina}, Meiramgul and {Bekdaulet}, Shukirgaliyev and {Zhumabay}, Nurman and {Manapbayeva}, Arailym},
        title = "{Infall Motions in the Hot Core Associated with the Hypercompact H II Region G345.0061+01.794 B}",
      journal = {\apj},
         year = 2024,
        month = may,
       volume = {967},
       number = {1},
          eid = {15},
        pages = {15},
          doi = {10.3847/1538-4357/ad3e7b},
archivePrefix = {arXiv},
       eprint = {2307.07459},
 primaryClass = {astro-ph.GA},
       adsurl = {https://ui.adsabs.harvard.edu/abs/2024ApJ...967...15K}
}

@ARTICLE{Keto2003,
       author = {{Keto}, Eric},
        title = "{The Formation of Massive Stars by Accretion through Trapped Hypercompact H II Regions}",
      journal = {\apj},
         year = 2003,
        month = dec,
       volume = {599},
       number = {2},
        pages = {1196-1206},
          doi = {10.1086/379545},
archivePrefix = {arXiv},
       eprint = {astro-ph/0309131},
 primaryClass = {astro-ph},
       adsurl = {https://ui.adsabs.harvard.edu/abs/2003ApJ...599.1196K}
}

@ARTICLE{Peters2010a,
       author = {{Peters}, Thomas and {Banerjee}, Robi and {Klessen}, Ralf S. and {Mac Low}, Mordecai-Mark and {Galv{\'a}n-Madrid}, Roberto and {Keto}, Eric R.},
        title = "{H II Regions: Witnesses to Massive Star Formation}",
      journal = {\apj},
         year = 2010,
        month = mar,
       volume = {711},
       number = {2},
        pages = {1017-1028},
          doi = {10.1088/0004-637X/711/2/1017},
archivePrefix = {arXiv},
       eprint = {1001.2470},
 primaryClass = {astro-ph.SR},
       adsurl = {https://ui.adsabs.harvard.edu/abs/2010ApJ...711.1017P}
}

@ARTICLE{Peters2010b,
       author = {{Peters}, Thomas and {Mac Low}, Mordecai-Mark and {Banerjee}, Robi and {Klessen}, Ralf S. and {Dullemond}, Cornelis P.},
        title = "{Understanding Spatial and Spectral Morphologies of Ultracompact H II Regions}",
      journal = {\apj},
         year = 2010,
        month = aug,
       volume = {719},
       number = {1},
        pages = {831-843},
          doi = {10.1088/0004-637X/719/1/831},
archivePrefix = {arXiv},
       eprint = {1003.4998},
 primaryClass = {astro-ph.GA},
       adsurl = {https://ui.adsabs.harvard.edu/abs/2010ApJ...719..831P}
}

@ARTICLE{DePree2014,
       author = {{De Pree}, C.~G. and {Peters}, T. and {Mac Low}, M. -M. and {Wilner}, D.~J. and {Goss}, W.~M. and {Galv{\'a}n-Madrid}, R. and {Keto}, E.~R. and {Klessen}, R.~S. and {Monsrud}, A.},
        title = "{Flickering of 1.3 cm Sources in Sgr B2: Toward a Solution to the Ultracompact H II Region Lifetime Problem}",
      journal = {\apjl},
         year = 2014,
        month = feb,
       volume = {781},
       number = {2},
          eid = {L36},
        pages = {L36},
          doi = {10.1088/2041-8205/781/2/L36},
       adsurl = {https://ui.adsabs.harvard.edu/abs/2014ApJ...781L..36D}
}

@ARTICLE{Galvan-Madrid2008,
       author = {{Galv{\'a}n-Madrid}, Roberto and {Rodr{\'\i}guez}, Luis F. and {Ho}, Paul T.~P. and {Keto}, Eric},
        title = "{Time Variation in G24.78+0.08 A1: Evidence for an Accreting Hypercompact H II Region?}",
      journal = {\apjl},
         year = 2008,
        month = feb,
       volume = {674},
       number = {1},
        pages = {L33},
          doi = {10.1086/528957},
archivePrefix = {arXiv},
       eprint = {0801.1298},
 primaryClass = {astro-ph},
       adsurl = {https://ui.adsabs.harvard.edu/abs/2008ApJ...674L..33G}
}

@ARTICLE{Yang2025,
       author = {{Yang}, A.~Y. and {Thompson}, M.~A. and {Urquhart}, J.~S. and {Brunthaler}, A. and {Menten}, K.~M. and {Gong}, Y. and {Tsai}, Chao-Wei and {Patel}, A.~L. and {Li}, D. and {Cotton}, W.~D.},
        title = "{Dynamic massive star formation: Radio flux variability in UC H II regions}",
      journal = {\aap},
         year = 2025,
        month = feb,
       volume = {694},
          eid = {A26},
        pages = {A26},
          doi = {10.1051/0004-6361/202452078},
archivePrefix = {arXiv},
       eprint = {2410.17117},
 primaryClass = {astro-ph.GA},
       adsurl = {https://ui.adsabs.harvard.edu/abs/2025A&A...694A..26Y}
}

@ARTICLE{Li2023,
       author = {{Li}, Shanghuo and {Sanhueza}, Patricio and {Zhang}, Qizhou and {Guido}, Garay and {Sabatini}, Giovanni and {Morii}, Kaho and {Lu}, Xing and {Tafoya}, Daniel and {Nakamura}, Fumitaka and {Izumi}, Natsuko and {Tatematsu}, Ken'ichi and {Li}, Fei},
        title = "{The ALMA Survey of 70 {\ensuremath{\mu}}m Dark High-mass Clumps in Early Stages (ASHES). VIII. Dynamics of Embedded Dense Cores}",
      journal = {\apj},
         year = 2023,
        month = jun,
       volume = {949},
       number = {2},
          eid = {109},
        pages = {109},
          doi = {10.3847/1538-4357/acc58f},
archivePrefix = {arXiv},
       eprint = {2304.01718},
 primaryClass = {astro-ph.GA},
       adsurl = {https://ui.adsabs.harvard.edu/abs/2023ApJ...949..109L}
}

@ARTICLE{Keto2008,
       author = {{Keto}, Eric and {Zhang}, Qizhou and {Kurtz}, Stanley},
        title = "{The Early Evolution of Massive Stars: Radio Recombination Line Spectra}",
      journal = {\apj},
         year = 2008,
        month = jan,
       volume = {672},
       number = {1},
        pages = {423-432},
          doi = {10.1086/522570},
archivePrefix = {arXiv},
       eprint = {0708.3388},
 primaryClass = {astro-ph},
       adsurl = {https://ui.adsabs.harvard.edu/abs/2008ApJ...672..423K}
}

@ARTICLE{Peters2012,
       author = {{Peters}, Thomas and {Longmore}, Steven N. and {Dullemond}, Cornelis P.},
        title = "{Understanding hydrogen recombination line observations with ALMA and EVLA}",
      journal = {\mnras},
         year = 2012,
        month = sep,
       volume = {425},
       number = {3},
        pages = {2352-2368},
          doi = {10.1111/j.1365-2966.2012.21676.x},
archivePrefix = {arXiv},
       eprint = {1206.7041},
 primaryClass = {astro-ph.GA},
       adsurl = {https://ui.adsabs.harvard.edu/abs/2012MNRAS.425.2352P}
}

@ARTICLE{Tenorio1979,
       author = {{Tenorio-Tagle}, G.},
        title = "{The gas dynamics of H II regions. I. The champagne model.}",
      journal = {\aap},
         year = 1979,
        month = jan,
       volume = {71},
        pages = {59-65},
       adsurl = {https://ui.adsabs.harvard.edu/abs/1979A&A....71...59T}
}

@ARTICLE{vanBuren1990,
       author = {{van Buren}, Dave and {Mac Low}, Mordecai-Mark and {Wood}, Douglas O.~S. and {Churchwell}, Ed},
        title = "{Cometary Compact H II Regions Are Stellar-Wind Bow Shocks}",
      journal = {\apj},
         year = 1990,
        month = apr,
       volume = {353},
        pages = {570},
          doi = {10.1086/168645},
       adsurl = {https://ui.adsabs.harvard.edu/abs/1990ApJ...353..570V}
}

@ARTICLE{Kim2017,
       author = {{Kim}, W. -J. and {Wyrowski}, F. and {Urquhart}, J.~S. and {Menten}, K.~M. and {Csengeri}, T.},
        title = "{ATLASGAL-selected massive clumps in the inner Galaxy. IV. Millimeter hydrogen recombination lines from associated H II regions}",
      journal = {\aap},
         year = 2017,
        month = jun,
       volume = {602},
          eid = {A37},
        pages = {A37},
          doi = {10.1051/0004-6361/201629764},
archivePrefix = {arXiv},
       eprint = {1702.02062},
 primaryClass = {astro-ph.GA},
       adsurl = {https://ui.adsabs.harvard.edu/abs/2017A&A...602A..37K}
}

@ARTICLE{Zapata2023,
       author = {{Zapata}, Luis A. and {Fern{\'a}ndez-L{\'o}pez}, Manuel and {Leurini}, Silvia and {Guzm{\'a}n Ccolque}, Estrella and {Skretas}, I.~M. and {Rodr{\'\i}guez}, Luis F. and {Palau},, Aina and {Menten}, Karl M. and {Wyrowski}, Friedrich},
        title = "{One, Two, Three ... An Explosive Outflow in IRAS 12326-6245 Revealed by ALMA}",
      journal = {\apjl},
         year = 2023,
        month = oct,
       volume = {956},
       number = {2},
          eid = {L35},
        pages = {L35},
          doi = {10.3847/2041-8213/acfe71},
archivePrefix = {arXiv},
       eprint = {2309.11386},
 primaryClass = {astro-ph.GA},
       adsurl = {https://ui.adsabs.harvard.edu/abs/2023ApJ...956L..35Z}
}

@ARTICLE{Surcis2019,
       author = {{Surcis}, G. and {Vlemmings}, W.~H.~T. and {van Langevelde}, H.~J. and {Hutawarakorn Kramer}, B. and {Bartkiewicz}, A.},
        title = "{EVN observations of 6.7 GHz methanol maser polarization in massive star-forming regions. IV. Magnetic field strength limits and structure for seven additional sources}",
      journal = {\aap},
         year = 2019,
        month = mar,
       volume = {623},
          eid = {A130},
        pages = {A130},
          doi = {10.1051/0004-6361/201834578},
archivePrefix = {arXiv},
       eprint = {1902.08210},
 primaryClass = {astro-ph.SR},
       adsurl = {https://ui.adsabs.harvard.edu/abs/2019A&A...623A.130S}
}

@ARTICLE{Kavak2021,
       author = {{Kavak}, {\"U}. and {S{\'a}nchez-Monge}, {\'A}. and {L{\'o}pez-Sepulcre}, A. and {Cesaroni}, R. and {van der Tak}, F.~F.~S. and {Moscadelli}, L. and {Beltr{\'a}n}, M.~T. and {Schilke}, P.},
        title = "{Search for radio jets from massive young stellar objects. Association of radio jets with H$_{2}$O and CH$_{3}$OH masers}",
      journal = {\aap},
         year = 2021,
        month = jan,
       volume = {645},
          eid = {A29},
        pages = {A29},
          doi = {10.1051/0004-6361/202037652},
archivePrefix = {arXiv},
       eprint = {2011.14729},
 primaryClass = {astro-ph.GA},
       adsurl = {https://ui.adsabs.harvard.edu/abs/2021A&A...645A..29K}
}

@ARTICLE{Wu2019,
       author = {{Wu}, Y.~W. and {Reid}, M.~J. and {Sakai}, N. and {Dame}, T.~M. and {Menten}, K.~M. and {Brunthaler}, A. and {Xu}, Y. and {Li}, J.~J. and {Ho}, B. and {Zhang}, B. and {Rygl}, K.~L.~J. and {Zheng}, X.~W.},
        title = "{Trigonometric Parallaxes of Star-forming Regions beyond the Tangent Point of the Sagittarius Spiral Arm}",
      journal = {\apj},
         year = 2019,
        month = mar,
       volume = {874},
       number = {1},
          eid = {94},
        pages = {94},
          doi = {10.3847/1538-4357/ab001a},
archivePrefix = {arXiv},
       eprint = {1901.09313},
 primaryClass = {astro-ph.GA},
       adsurl = {https://ui.adsabs.harvard.edu/abs/2019ApJ...874...94W}
}

@ARTICLE{Kim2019,
       author = {{Kim}, Won-Ju and {Kim}, Kee-Tae and {Kim}, Kwang-Tae},
        title = "{Simultaneous 22 GHz Water and 44 GHz Methanol Maser Survey of Ultracompact H II Regions}",
      journal = {\apjs},
         year = 2019,
        month = sep,
       volume = {244},
       number = {1},
          eid = {2},
        pages = {2},
          doi = {10.3847/1538-4365/ab2fc9},
archivePrefix = {arXiv},
       eprint = {1907.11593},
 primaryClass = {astro-ph.SR},
       adsurl = {https://ui.adsabs.harvard.edu/abs/2019ApJS..244....2K}
}

@ARTICLE{Fish2005,
       author = {{Fish}, Vincent L. and {Reid}, Mark J. and {Argon}, Alice L. and {Zheng}, Xing-Wu},
        title = "{Full-Polarization Observations of OH Masers in Massive Star-forming Regions. I. Data}",
      journal = {\apjs},
         year = 2005,
        month = sep,
       volume = {160},
       number = {1},
        pages = {220-271},
          doi = {10.1086/431669},
archivePrefix = {arXiv},
       eprint = {astro-ph/0505148},
 primaryClass = {astro-ph},
       adsurl = {https://ui.adsabs.harvard.edu/abs/2005ApJS..160..220F}
}

@ARTICLE{Baudry1997,
       author = {{Baudry}, A. and {Desmurs}, J.~F. and {Wilson}, T.~L. and {Cohen}, R.~J.},
        title = "{A survey of star-forming regions in the 5 CM lines of OH.}",
      journal = {\aap},
         year = 1997,
        month = sep,
       volume = {325},
        pages = {255-268},
       adsurl = {https://ui.adsabs.harvard.edu/abs/1997A&A...325..255B}
}

@ARTICLE{Winnberg1975,
       author = {{Winnberg}, A. and {Nguyen-Quang-Rieu} and {Johansson}, L.~E.~B. and {Goss}, W.~M.},
        title = "{Positions of OH Sources Discovered at 1612 MHz}",
      journal = {\aap},
         year = 1975,
        month = jan,
       volume = {38},
        pages = {145},
       adsurl = {https://ui.adsabs.harvard.edu/abs/1975A&A....38..145W}
}

@ARTICLE{Honma2005,
       author = {{Honma}, Mareki and {Bushimata}, Takeshi and {Choi}, Yoon Kyung and {Fujii}, Takahiro and {Hirota}, Tomoya and {Horiai}, Koji and {Imai}, Hiroshi and {Inomata}, Noritomo and {Ishitsuka}, Jose and {Iwadate}, Kenzaburo and {Jike}, Takaaki and {Kameya}, Osamu and {Kamohara}, Ryuichi and {Kan-Ya}, Yukitoshi and {Kawaguchi}, Noriyuki and {Kijima}, Masachika and {Kobayashi}, Hideyuki and {Kuji}, Seisuke and {Kurayama}, Tomoharu and {Manabe}, Seiji and {Miyaji}, Takeshi and {Nakagawa}, Akiharu and {Nakashima}, Kouichirou and {Oh}, Chung Sik and {Omodaka}, Toshihiro and {Oyama}, Tomoaki and {Rioja}, Maria and {Sakai}, Satoshi and {Sato}, Katsuhisa and {Sasao}, Tetsuo and {Shibata}, Katsunori M. and {Shimizu}, Rie and {Sora}, Kasumi and {Suda}, Hiroshi and {Tamura}, Yoshiaki and {Yamashita}, Kazuyoshi},
        title = "{Multi-Epoch VERA Observations of H$_{2}$O Masers in OH 43.8-0.1}",
      journal = {\pasj},
         year = 2005,
        month = aug,
       volume = {57},
        pages = {595-603},
          doi = {10.1093/pasj/57.4.595},
       adsurl = {https://ui.adsabs.harvard.edu/abs/2005PASJ...57..595H}
}

@ARTICLE{Urquhart2022,
       author = {{Urquhart}, J.~S. and {Wells}, M.~R.~A. and {Pillai}, T. and {Leurini}, S. and {Giannetti}, A. and {Moore}, T.~J.~T. and {Thompson}, M.~A. and {Figura}, C. and {Colombo}, D. and {Yang}, A.~Y. and {K{\"o}nig}, C. and {Wyrowski}, F. and {Menten}, K.~M. and {Rigby}, A.~J. and {Eden}, D.~J. and {Ragan}, S.~E.},
        title = "{ATLASGAL - evolutionary trends in high-mass star formation}",
      journal = {\mnras},
         year = 2022,
        month = mar,
       volume = {510},
       number = {3},
        pages = {3389-3407},
          doi = {10.1093/mnras/stab3511},
archivePrefix = {arXiv},
       eprint = {2111.12816},
 primaryClass = {astro-ph.GA},
       adsurl = {https://ui.adsabs.harvard.edu/abs/2022MNRAS.510.3389U}
}

@ARTICLE{Koch2015,
   author = {{Koch}, E.~W. and {Rosolowsky}, E.~W.},
    title = "{Filament identification through mathematical morphology}",
  journal = {\mnras},
archivePrefix = "arXiv",
   eprint = {1507.02289},
     year = 2015,
    month = oct,
   volume = 452,
    pages = {3435-3450},
      doi = {10.1093/mnras/stv1521},
   adsurl = {http://adsabs.harvard.edu/abs/2015MNRAS.452.3435K}
}

@ARTICLE{Kurtz1994,
       author = {{Kurtz}, S. and {Churchwell}, E. and {Wood}, D.~O.~S.},
        title = "{Ultracompact H II Regions. II. New High-Resolution Radio Images}",
      journal = {\apjs},
         year = 1994,
        month = apr,
       volume = {91},
        pages = {659},
          doi = {10.1086/191952},
       adsurl = {https://ui.adsabs.harvard.edu/abs/1994ApJS...91..659K}
}

@ARTICLE{Kalcheva2018,
       author = {{Kalcheva}, I.~E. and {Hoare}, M.~G. and {Urquhart}, J.~S. and {Kurtz}, S. and {Lumsden}, S.~L. and {Purcell}, C.~R. and {Zijlstra}, A.~A.},
        title = "{The coordinated radio and infrared survey for high-mass star formation. III. A catalogue of northern ultra-compact H II regions}",
      journal = {\aap},
         year = 2018,
        month = jul,
       volume = {615},
          eid = {A103},
        pages = {A103},
          doi = {10.1051/0004-6361/201832734},
archivePrefix = {arXiv},
       eprint = {1803.09334},
 primaryClass = {astro-ph.GA},
       adsurl = {https://ui.adsabs.harvard.edu/abs/2018A&A...615A.103K}
}

@ARTICLE{XCLASS2017,
      author = {{M{\"o}ller}, T. and {Endres}, C. and {Schilke}, P.},
      title = "{eXtended CASA Line Analysis Software Suite (XCLASS)}",
      journal = {\aap},
         year = 2017,
      month = feb,
      volume = {598},
         eid = {A7},
      pages = {A7},
         doi = {10.1051/0004-6361/201527203},
archivePrefix = {arXiv},
      eprint = {1508.04114},
primaryClass = {astro-ph.IM},
      adsurl = {https://ui.adsabs.harvard.edu/abs/2017A&A...598A...7M}
}

@ARTICLE{Sewilo2004,
       author = {{Sewilo}, M. and {Churchwell}, E. and {Kurtz}, S. and {Goss}, W.~M. and {Hofner}, P.},
        title = "{Broad Radio Recombination Lines from Hypercompact H II Regions}",
      journal = {\apj},
         year = 2004,
        month = apr,
       volume = {605},
       number = {1},
        pages = {285-299},
          doi = {10.1086/382268},
       adsurl = {https://ui.adsabs.harvard.edu/abs/2004ApJ...605..285S}
}

@ARTICLE{Vacca1996,
       author = {{Vacca}, William D. and {Garmany}, Catharine D. and {Shull}, J. Michael},
        title = "{The Lyman-Continuum Fluxes and Stellar Parameters of O and Early B-Type Stars}",
      journal = {\apj},
         year = 1996,
        month = apr,
       volume = {460},
        pages = {914},
          doi = {10.1086/177020},
       adsurl = {https://ui.adsabs.harvard.edu/abs/1996ApJ...460..914V}
}

@ARTICLE{Klaassen2018,
       author = {{Klaassen}, P.~D. and {Johnston}, K.~G. and {Urquhart}, J.~S. and {Mottram}, J.~C. and {Peters}, T. and {Kuiper}, R. and {Beuther}, H. and {van der Tak}, F.~F.~S. and {Goddi}, C.},
        title = "{The evolution of young HII regions. I. Continuum emission and internal dynamics}",
      journal = {\aap},
         year = 2018,
        month = apr,
       volume = {611},
          eid = {A99},
        pages = {A99},
          doi = {10.1051/0004-6361/201731727},
       adsurl = {https://ui.adsabs.harvard.edu/abs/2018A&A...611A..99K}
}

@BOOK{Wilson2013,
       author = {{Wilson}, Thomas L. and {Rohlfs}, Kristen and {H{\"u}ttemeister}, Susanne},
        title = "{Tools of Radio Astronomy}",
         year = 2013,
          doi = {10.1007/978-3-642-39950-3},
       adsurl = {https://ui.adsabs.harvard.edu/abs/2013tra..book.....W},
      publisher = "{Springer}"

}

@BOOK{Condon2016,
       author = {{Condon}, James J. and {Ransom}, Scott M.},
        title = "{Essential Radio Astronomy}",
         year = 2016,
       adsurl = {https://ui.adsabs.harvard.edu/abs/2016era..book.....C},
      publisher = "{Princeton University Press}"
}

@misc{Dullemond2012,
       author = {{Dullemond}, C.~P. and {Juhasz}, A. and {Pohl}, A. and {Sereshti}, F. and {Shetty}, R. and {Peters}, T. and {Commercon}, B. and {Flock}, M.},
        title = "{RADMC-3D: A multi-purpose radiative transfer tool}",
 howpublished = {Astrophysics Source Code Library, record ascl:1202.015},
         year = 2012,
        month = feb,
          eid = {ascl:1202.015},
       adsurl = {https://ui.adsabs.harvard.edu/abs/2012ascl.soft02015D}
}

@ARTICLE{Fujii2011,
       author = {{Fujii}, Michiko S. and {Portegies Zwart}, Simon},
        title = "{The Origin of OB Runaway Stars}",
      journal = {Science},
         year = 2011,
        month = dec,
       volume = {334},
       number = {6061},
        pages = {1380},
          doi = {10.1126/science.1211927},
archivePrefix = {arXiv},
       eprint = {1111.3644},
 primaryClass = {astro-ph.GA},
       adsurl = {https://ui.adsabs.harvard.edu/abs/2011Sci...334.1380F}
}

@ARTICLE{vanBuren1992,
       author = {{van Buren}, Dave and {Mac Low}, Mordecai-Mark},
        title = "{Bow Shock Models for the Velocity Structure of Ultracompact H II Regions}",
      journal = {\apj},
         year = 1992,
        month = aug,
       volume = {394},
        pages = {534},
          doi = {10.1086/171604},
       adsurl = {https://ui.adsabs.harvard.edu/abs/1992ApJ...394..534V}
}

@ARTICLE{Zhu2008,
       author = {{Zhu}, Qing-Feng and {Lacy}, John H. and {Jaffe}, Daniel T. and {Richter}, Matthew J. and {Greathouse}, Thomas K.},
        title = "{[Ne II] Observations of Gas Motions in Compact and Ultracompact H II Regions}",
      journal = {\apjs},
         year = 2008,
        month = aug,
       volume = {177},
       number = {2},
        pages = {584-612},
          doi = {10.1086/588731},
archivePrefix = {arXiv},
       eprint = {0804.0439},
 primaryClass = {astro-ph},
       adsurl = {https://ui.adsabs.harvard.edu/abs/2008ApJS..177..584Z}
}

@ARTICLE{Wilkin1996,
       author = {{Wilkin}, Francis P.},
        title = "{Exact Analytic Solutions for Stellar Wind Bow Shocks}",
      journal = {\apjl},
         year = 1996,
        month = mar,
       volume = {459},
        pages = {L31},
          doi = {10.1086/309939},
       adsurl = {https://ui.adsabs.harvard.edu/abs/1996ApJ...459L..31W}
}

@ARTICLE{Zhu2015,
       author = {{Zhu}, Feng-Yao and {Zhu}, Qing-Feng and {Li}, Juan and {Zhang}, Jiang-Shui and {Wang}, Jun-Zhi},
        title = "{The Hydrodynamical Models of the Cometary Compact HII Region}",
      journal = {\apj},
         year = 2015,
        month = oct,
       volume = {812},
       number = {1},
          eid = {87},
        pages = {87},
          doi = {10.1088/0004-637X/812/1/87},
archivePrefix = {arXiv},
       eprint = {1510.03183},
 primaryClass = {astro-ph.SR},
       adsurl = {https://ui.adsabs.harvard.edu/abs/2015ApJ...812...87Z}
}

@ARTICLE{Tenorio1979b,
       author = {{Tenorio-Tagle}, G. and {Yorke}, H.~W. and {Bodenheimer}, P.},
        title = "{The gas dynamics of HII regions III. The components of the galactic extended low density HII region.}",
      journal = {\aap},
         year = 1979,
        month = nov,
       volume = {145},
        pages = {110},
       adsurl = {https://ui.adsabs.harvard.edu/abs/1979A&A....80..110T}
}

@ARTICLE{Bodenheimer1979,
       author = {{Bodenheimer}, P. and {Tenorio-Tagle}, G. and {Yorke}, H.~W.},
        title = "{The gas dynamics of H II regions. II. Two-dimensional axisymmetric calculations.}",
      journal = {\apj},
         year = 1979,
        month = oct,
       volume = {233},
        pages = {85-96},
          doi = {10.1086/157368},
       adsurl = {https://ui.adsabs.harvard.edu/abs/1979ApJ...233...85B}
}

@ARTICLE{Arthur2006,
       author = {{Arthur}, S. Jane and {Hoare}, M.~G.},
        title = "{Hydrodynamics of Cometary Compact H II Regions}",
      journal = {\apjs},
         year = 2006,
        month = jul,
       volume = {165},
       number = {1},
        pages = {283-306},
          doi = {10.1086/503899},
archivePrefix = {arXiv},
       eprint = {astro-ph/0511035},
 primaryClass = {astro-ph},
       adsurl = {https://ui.adsabs.harvard.edu/abs/2006ApJS..165..283A}
}

@ARTICLE{Elitzur1978,
       author = {{Elitzur}, M. and {de Jong}, T.},
        title = "{A model for the maser sources associated with H II regions.}",
      journal = {\aap},
         year = 1978,
        month = jul,
       volume = {67},
       number = {3},
        pages = {323-332},
       adsurl = {https://ui.adsabs.harvard.edu/abs/1978A&A....67..323E}
}

@ARTICLE{Fish2007,
       author = {{Fish}, Vincent L. and {Reid}, Mark J.},
        title = "{Proper Motions of OH Masers and Magnetic Fields in Massive Star-forming Regions}",
      journal = {\apj},
         year = 2007,
        month = dec,
       volume = {670},
       number = {2},
        pages = {1159-1172},
          doi = {10.1086/522329},
archivePrefix = {arXiv},
       eprint = {0708.1186},
 primaryClass = {astro-ph},
       adsurl = {https://ui.adsabs.harvard.edu/abs/2007ApJ...670.1159F}
}

@ARTICLE{Gasiprong2002,
       author = {{Gasiprong}, N. and {Cohen}, R.~J. and {Hutawarakorn}, B.},
        title = "{OH masers and magnetic fields near the cometary HII region G34.3+0.2}",
      journal = {\mnras},
         year = 2002,
        month = oct,
       volume = {336},
       number = {1},
        pages = {47-54},
          doi = {10.1046/j.1365-8711.2002.05703.x},
       adsurl = {https://ui.adsabs.harvard.edu/abs/2002MNRAS.336...47G}
}

@ARTICLE{Keto1995,
       author = {{Keto}, E.~R. and {Welch}, W.~J. and {Reid}, M.~J. and {Ho}, P.~T.~P.},
        title = "{Line Broadening in the W3(OH) Champagne Flow}",
      journal = {\apj},
         year = 1995,
        month = may,
       volume = {444},
        pages = {765},
          doi = {10.1086/175649},
       adsurl = {https://ui.adsabs.harvard.edu/abs/1995ApJ...444..765K}
}

@ARTICLE{Wood1989,
       author = {{Wood}, Douglas O.~S. and {Churchwell}, Ed},
        title = "{The Morphologies and Physical Properties of Ultracompact H II Regions}",
      journal = {\apjs},
         year = 1989,
        month = apr,
       volume = {69},
        pages = {831},
          doi = {10.1086/191329},
       adsurl = {https://ui.adsabs.harvard.edu/abs/1989ApJS...69..831W}
}

@ARTICLE{Zhang2024,
       author = {{Zhang}, Siju and {Liu}, Tie and {Wang}, Ke and {Zavagno}, Annie and {Garay}, Guido and {Liu}, Hongli and {Xu}, Fengwei and {Liu}, Xunchuan and {Sanhueza}, Patricio and {Soam}, Archana and {Zhou}, Jian-wen and {Li}, Shanghuo and {Goldsmith}, Paul F. and {Zhang}, Yong and {Chibueze}, James O. and {Lee}, Chang Won and {Hwang}, Jihye and {Bronfman}, Leonardo and {Dewangan}, Lokesh K.},
        title = "{ATOMS: ALMA three-millimeter observations of massive star-forming regions - XVIII. On the origin and evolution of dense gas fragments in molecular shells of compact H II regions}",
      journal = {\mnras},
         year = 2024,
        month = dec,
       volume = {535},
       number = {2},
        pages = {1364-1386},
          doi = {10.1093/mnras/stae2415},
archivePrefix = {arXiv},
       eprint = {2410.17455},
 primaryClass = {astro-ph.GA},
       adsurl = {https://ui.adsabs.harvard.edu/abs/2024MNRAS.535.1364Z}
}

@ARTICLE{Hoare2005,
       author = {{Hoare}, M.~G.},
        title = "{Ultra-Compact H II Regions}",
      journal = {\apss},
         year = 2005,
        month = jan,
       volume = {295},
       number = {1-2},
        pages = {203-215},
          doi = {10.1007/s10509-005-3690-1},
       adsurl = {https://ui.adsabs.harvard.edu/abs/2005Ap&SS.295..203H}
}

@INPROCEEDINGS{Hoare2007,
       author = {{Hoare}, M.~G. and {Kurtz}, S.~E. and {Lizano}, S. and {Keto}, E. and {Hofner}, P.},
        title = "{Ultracompact Hii Regions and the Early Lives of Massive Stars}",
    booktitle = {Protostars and Planets V},
         year = 2007,
       editor = {{Reipurth}, Bo and {Jewitt}, David and {Keil}, Klaus},
        month = jan,
        pages = {181},
          doi = {10.48550/arXiv.astro-ph/0603560},
archivePrefix = {arXiv},
       eprint = {astro-ph/0603560},
 primaryClass = {astro-ph},
       adsurl = {https://ui.adsabs.harvard.edu/abs/2007prpl.conf..181H}
}

@ARTICLE{Churchwell2002,
       author = {{Churchwell}, Ed},
        title = "{Ultra-Compact HII Regions and Massive Star Formation}",
      journal = {\araa},
         year = 2002,
        month = jan,
       volume = {40},
        pages = {27-62},
          doi = {10.1146/annurev.astro.40.060401.093845},
       adsurl = {https://ui.adsabs.harvard.edu/abs/2002ARA&A..40...27C}
}

@ARTICLE{Garay1999,
       author = {{Garay}, Guido and {Lizano}, Susana},
        title = "{Massive Stars: Their Environment and Formation}",
      journal = {\pasp},
         year = 1999,
        month = sep,
       volume = {111},
       number = {763},
        pages = {1049-1087},
          doi = {10.1086/316416},
archivePrefix = {arXiv},
       eprint = {astro-ph/9907293},
 primaryClass = {astro-ph},
       adsurl = {https://ui.adsabs.harvard.edu/abs/1999PASP..111.1049G}
}

@ARTICLE{Geen2021,
       author = {{Geen}, Sam and {Bieri}, Rebekka and {Rosdahl}, Joakim and {de Koter}, Alex},
        title = "{The geometry and dynamical role of stellar wind bubbles in photoionized H II regions}",
      journal = {\mnras},
         year = 2021,
        month = feb,
       volume = {501},
       number = {1},
        pages = {1352-1369},
          doi = {10.1093/mnras/staa3705},
archivePrefix = {arXiv},
       eprint = {2009.08742},
 primaryClass = {astro-ph.GA},
       adsurl = {https://ui.adsabs.harvard.edu/abs/2021MNRAS.501.1352G}
}

@ARTICLE{Astropy2022,
       author = {{Astropy Collaboration} and {Price-Whelan}, Adrian M. and {Lim}, Pey Lian and {Earl}, Nicholas and {Starkman}, Nathaniel and {Bradley}, Larry and {Shupe}, David L. and {Patil}, Aarya A. and {Corrales}, Lia and {Brasseur}, C.~E. and {N{\"o}the}, Maximilian and {Donath}, Axel and {Tollerud}, Erik and {Morris}, Brett M. and {Ginsburg}, Adam and {Vaher}, Eero and {Weaver}, Benjamin A. and {Tocknell}, James and {Jamieson}, William and {van Kerkwijk}, Marten H. and {Robitaille}, Thomas P. and {Merry}, Bruce and {Bachetti}, Matteo and {G{\"u}nther}, H. Moritz and {Aldcroft}, Thomas L. and {Alvarado-Montes}, Jaime A. and {Archibald}, Anne M. and {B{\'o}di}, Attila and {Bapat}, Shreyas and {Barentsen}, Geert and {Baz{\'a}n}, Juanjo and {Biswas}, Manish and {Boquien}, M{\'e}d{\'e}ric and {Burke}, D.~J. and {Cara}, Daria and {Cara}, Mihai and {Conroy}, Kyle E. and {Conseil}, Simon and {Craig}, Matthew W. and {Cross}, Robert M. and {Cruz}, Kelle L. and {D'Eugenio}, Francesco and {Dencheva}, Nadia and {Devillepoix}, Hadrien A.~R. and {Dietrich}, J{\"o}rg P. and {Eigenbrot}, Arthur Davis and {Erben}, Thomas and {Ferreira}, Leonardo and {Foreman-Mackey}, Daniel and {Fox}, Ryan and {Freij}, Nabil and {Garg}, Suyog and {Geda}, Robel and {Glattly}, Lauren and {Gondhalekar}, Yash and {Gordon}, Karl D. and {Grant}, David and {Greenfield}, Perry and {Groener}, Austen M. and {Guest}, Steve and {Gurovich}, Sebastian and {Handberg}, Rasmus and {Hart}, Akeem and {Hatfield-Dodds}, Zac and {Homeier}, Derek and {Hosseinzadeh}, Griffin and {Jenness}, Tim and {Jones}, Craig K. and {Joseph}, Prajwel and {Kalmbach}, J. Bryce and {Karamehmetoglu}, Emir and {Ka{\l}uszy{\'n}ski}, Miko{\l}aj and {Kelley}, Michael S.~P. and {Kern}, Nicholas and {Kerzendorf}, Wolfgang E. and {Koch}, Eric W. and {Kulumani}, Shankar and {Lee}, Antony and {Ly}, Chun and {Ma}, Zhiyuan and {MacBride}, Conor and {Maljaars}, Jakob M. and {Muna}, Demitri and {Murphy}, N.~A. and {Norman}, Henrik and {O'Steen}, Richard and {Oman}, Kyle A. and {Pacifici}, Camilla and {Pascual}, Sergio and {Pascual-Granado}, J. and {Patil}, Rohit R. and {Perren}, Gabriel I. and {Pickering}, Timothy E. and {Rastogi}, Tanuj and {Roulston}, Benjamin R. and {Ryan}, Daniel F. and {Rykoff}, Eli S. and {Sabater}, Jose and {Sakurikar}, Parikshit and {Salgado}, Jes{\'u}s and {Sanghi}, Aniket and {Saunders}, Nicholas and {Savchenko}, Volodymyr and {Schwardt}, Ludwig and {Seifert-Eckert}, Michael and {Shih}, Albert Y. and {Jain}, Anany Shrey and {Shukla}, Gyanendra and {Sick}, Jonathan and {Simpson}, Chris and {Singanamalla}, Sudheesh and {Singer}, Leo P. and {Singhal}, Jaladh and {Sinha}, Manodeep and {Sip{\H{o}}cz}, Brigitta M. and {Spitler}, Lee R. and {Stansby}, David and {Streicher}, Ole and {{\v{S}}umak}, Jani and {Swinbank}, John D. and {Taranu}, Dan S. and {Tewary}, Nikita and {Tremblay}, Grant R. and {de Val-Borro}, Miguel and {Van Kooten}, Samuel J. and {Vasovi{\'c}}, Zlatan and {Verma}, Shresth and {de Miranda Cardoso}, Jos{\'e} Vin{\'\i}cius and {Williams}, Peter K.~G. and {Wilson}, Tom J. and {Winkel}, Benjamin and {Wood-Vasey}, W.~M. and {Xue}, Rui and {Yoachim}, Peter and {Zhang}, Chen and {Zonca}, Andrea and {Astropy Project Contributors}},
        title = "{The Astropy Project: Sustaining and Growing a Community-oriented Open-source Project and the Latest Major Release (v5.0) of the Core Package}",
      journal = {\apj},
         year = 2022,
        month = aug,
       volume = {935},
       number = {2},
          eid = {167},
        pages = {167},
          doi = {10.3847/1538-4357/ac7c74},
archivePrefix = {arXiv},
       eprint = {2206.14220},
 primaryClass = {astro-ph.IM},
       adsurl = {https://ui.adsabs.harvard.edu/abs/2022ApJ...935..167A}
}

@ARTICLE{CASA2022,
       author = {{CASA Team} and {Bean}, Ben and {Bhatnagar}, Sanjay and {Castro}, Sandra and {Donovan Meyer}, Jennifer and {Emonts}, Bjorn and {Garcia}, Enrique and {Garwood}, Robert and {Golap}, Kumar and {Gonzalez Villalba}, Justo and {Harris}, Pamela and {Hayashi}, Yohei and {Hoskins}, Josh and {Hsieh}, Mingyu and {Jagannathan}, Preshanth and {Kawasaki}, Wataru and {Keimpema}, Aard and {Kettenis}, Mark and {Lopez}, Jorge and {Marvil}, Joshua and {Masters}, Joseph and {McNichols}, Andrew and {Mehringer}, David and {Miel}, Renaud and {Moellenbrock}, George and {Montesino}, Federico and {Nakazato}, Takeshi and {Ott}, Juergen and {Petry}, Dirk and {Pokorny}, Martin and {Raba}, Ryan and {Rau}, Urvashi and {Schiebel}, Darrell and {Schweighart}, Neal and {Sekhar}, Srikrishna and {Shimada}, Kazuhiko and {Small}, Des and {Steeb}, Jan-Willem and {Sugimoto}, Kanako and {Suoranta}, Ville and {Tsutsumi}, Takahiro and {van Bemmel}, Ilse M. and {Verkouter}, Marjolein and {Wells}, Akeem and {Xiong}, Wei and {Szomoru}, Arpad and {Griffith}, Morgan and {Glendenning}, Brian and {Kern}, Jeff},
        title = "{CASA, the Common Astronomy Software Applications for Radio Astronomy}",
      journal = {\pasp},
         year = 2022,
        month = nov,
       volume = {134},
       number = {1041},
          eid = {114501},
        pages = {114501},
          doi = {10.1088/1538-3873/ac9642},
archivePrefix = {arXiv},
       eprint = {2210.02276},
 primaryClass = {astro-ph.IM},
       adsurl = {https://ui.adsabs.harvard.edu/abs/2022PASP..134k4501C}
}

@ARTICLE{Ossenkopf1994,
       author = {{Ossenkopf}, V. and {Henning}, Th.},
        title = "{Dust opacities for protostellar cores.}",
      journal = {\aap},
         year = 1994,
        month = nov,
       volume = {291},
        pages = {943-959},
       adsurl = {https://ui.adsabs.harvard.edu/abs/1994A&A...291..943O}
}

@ARTICLE{Zhang2023,
       author = {{Zhang}, C. and {Zhu}, Feng-Yao and {Liu}, Tie and {Ren}, Z. -Y. and {Liu}, H. -L. and {Wang}, Ke and {Wu}, J. -W. and {Zhang}, Y. and {Zhou}, J. -W. and {Tatematsu}, K. and {Garay}, Guido and {Tej}, Anandmayee and {Li}, Shanghuo and {Xu}, W.~F. and {Lee}, Chang Won and {Bronfman}, Leonardo and {Soam}, Archana and {Li}, D.},
        title = "{ATOMS: ALMA three-millimetre observations of massive star-forming regions - XIV. Properties of resolved ultra-compact H II regions}",
      journal = {\mnras},
         year = 2023,
        month = apr,
       volume = {520},
       number = {3},
        pages = {3245-3258},
          doi = {10.1093/mnras/stad190},
archivePrefix = {arXiv},
       eprint = {2301.01876},
 primaryClass = {astro-ph.GA},
       adsurl = {https://ui.adsabs.harvard.edu/abs/2023MNRAS.520.3245Z}
}

@ARTICLE{Menshchikov2021,
       author = {{Men'shchikov}, A.},
        title = "{Multiscale, multiwavelength extraction of sources and filaments using separation of the structural components: getsf}",
      journal = {\aap},
         year = 2021,
        month = may,
       volume = {649},
          eid = {A89},
        pages = {A89},
          doi = {10.1051/0004-6361/202039913},
archivePrefix = {arXiv},
       eprint = {2102.11565},
 primaryClass = {astro-ph.IM},
       adsurl = {https://ui.adsabs.harvard.edu/abs/2021A&A...649A..89M}
}

@ARTICLE{Suin2025,
       author = {{Suin}, P. and {Arzoumanian}, D. and {Zavagno}, A. and {Hennebelle}, P.},
        title = "{The role of magnetic field and stellar feedback in the evolution of filamentary structures in collapsing star-forming clouds}",
      journal = {\aap},
         year = 2025,
        month = jun,
       volume = {698},
          eid = {A119},
        pages = {A119},
          doi = {10.1051/0004-6361/202553795},
archivePrefix = {arXiv},
       eprint = {2505.02903},
 primaryClass = {astro-ph.GA},
       adsurl = {https://ui.adsabs.harvard.edu/abs/2025A&A...698A.119S}
}

@ARTICLE{Yang2019,
       author = {{Yang}, A.~Y. and {Thompson}, M.~A. and {Tian}, W.~W. and {Bihr}, S. and {Beuther}, H. and {Hindson}, L.},
        title = "{A search for hypercompact H II regions in the Galactic Plane}",
      journal = {\mnras},
         year = 2019,
        month = jan,
       volume = {482},
       number = {2},
        pages = {2681-2696},
          doi = {10.1093/mnras/sty2811},
archivePrefix = {arXiv},
       eprint = {1809.00404},
 primaryClass = {astro-ph.GA},
       adsurl = {https://ui.adsabs.harvard.edu/abs/2019MNRAS.482.2681Y}
}

@ARTICLE{Yang2021,
       author = {{Yang}, A.~Y. and {Urquhart}, J.~S. and {Thompson}, M.~A. and {Menten}, K.~M. and {Wyrowski}, F. and {Brunthaler}, A. and {Tian}, W.~W. and {Rugel}, M. and {Yang}, X.~L. and {Yao}, S. and {Mutale}, M.},
        title = "{A population of hypercompact H II regions identified from young H II regions}",
      journal = {\aap},
         year = 2021,
        month = jan,
       volume = {645},
          eid = {A110},
        pages = {A110},
          doi = {10.1051/0004-6361/202038608},
archivePrefix = {arXiv},
       eprint = {2011.07620},
 primaryClass = {astro-ph.GA},
       adsurl = {https://ui.adsabs.harvard.edu/abs/2021A&A...645A.110Y}
}

@ARTICLE{Rodriguez2020,
       author = {{Rodr{\'\i}guez}, Luis F. and {Galv{\'a}n-Madrid}, Roberto and {Sanchez-Bermudez}, Joel and {De Pree}, Christopher G.},
        title = "{A Massive Young Runaway Star in W49 North}",
      journal = {\apj},
         year = 2020,
        month = feb,
       volume = {890},
       number = {2},
          eid = {165},
        pages = {165},
          doi = {10.3847/1538-4357/ab7011},
archivePrefix = {arXiv},
       eprint = {2002.02894},
 primaryClass = {astro-ph.SR},
       adsurl = {https://ui.adsabs.harvard.edu/abs/2020ApJ...890..165R}
}

@ARTICLE{Reed2005,
       author = {{Reed}, B. Cameron},
        title = "{New Estimates of the Solar-Neighborhood Massive Star Birthrate and the Galactic Supernova Rate}",
      journal = {\aj},
         year = 2005,
        month = oct,
       volume = {130},
       number = {4},
        pages = {1652-1657},
          doi = {10.1086/444474},
archivePrefix = {arXiv},
       eprint = {astro-ph/0506708},
 primaryClass = {astro-ph},
       adsurl = {https://ui.adsabs.harvard.edu/abs/2005AJ....130.1652R}
}

@ARTICLE{Robitaille2010,
       author = {{Robitaille}, Thomas P. and {Whitney}, Barbara A.},
        title = "{The Present-Day Star Formation Rate of the Milky Way Determined from Spitzer-Detected Young Stellar Objects}",
      journal = {\apjl},
         year = 2010,
        month = feb,
       volume = {710},
       number = {1},
        pages = {L11-L15},
          doi = {10.1088/2041-8205/710/1/L11},
archivePrefix = {arXiv},
       eprint = {1001.3672},
 primaryClass = {astro-ph.GA},
       adsurl = {https://ui.adsabs.harvard.edu/abs/2010ApJ...710L..11R}
}

@ARTICLE{Chomiuk2011,
       author = {{Chomiuk}, Laura and {Povich}, Matthew S.},
        title = "{Toward a Unification of Star Formation Rate Determinations in the Milky Way and Other Galaxies}",
      journal = {\aj},
         year = 2011,
        month = dec,
       volume = {142},
       number = {6},
          eid = {197},
        pages = {197},
          doi = {10.1088/0004-6256/142/6/197},
archivePrefix = {arXiv},
       eprint = {1110.4105},
 primaryClass = {astro-ph.GA},
       adsurl = {https://ui.adsabs.harvard.edu/abs/2011AJ....142..197C}
}

@ARTICLE{Beuther2025,
       author = {{Beuther}, H. and {Kuiper}, R. and {Tafalla}, M.},
        title = "{Star formation from low to high mass: A comparative view}",
      journal = {arXiv e-prints},
         year = 2025,
        month = jan,
          eid = {arXiv:2501.16866},
        pages = {arXiv:2501.16866},
          doi = {10.48550/arXiv.2501.16866},
archivePrefix = {arXiv},
       eprint = {2501.16866},
 primaryClass = {astro-ph.GA},
       adsurl = {https://ui.adsabs.harvard.edu/abs/2025arXiv250116866B}
}

@ARTICLE{Olguin2021,
       author = {{Olguin}, Fernando A. and {Sanhueza}, Patricio and {Guzm{\'a}n}, Andr{\'e}s E. and {Lu}, Xing and {Saigo}, Kazuya and {Zhang}, Qizhou and {Silva}, Andrea and {Chen}, Huei-Ru Vivien and {Li}, Shanghuo and {Ohashi}, Satoshi and {Nakamura}, Fumitaka and {Sakai}, Takeshi and {Wu}, Benjamin},
        title = "{Digging into the Interior of Hot Cores with ALMA (DIHCA). I. Dissecting the High-mass Star-forming Core G335.579-0.292 MM1}",
      journal = {\apj},
         year = 2021,
        month = mar,
       volume = {909},
       number = {2},
          eid = {199},
        pages = {199},
          doi = {10.3847/1538-4357/abde3f},
archivePrefix = {arXiv},
       eprint = {2101.08284},
 primaryClass = {astro-ph.GA},
       adsurl = {https://ui.adsabs.harvard.edu/abs/2021ApJ...909..199O}
}

@ARTICLE{Olguin2022,
       author = {{Olguin}, Fernando A. and {Sanhueza}, Patricio and {Ginsburg}, Adam and {Chen}, Huei-Ru Vivien and {Zhang}, Qizhou and {Li}, Shanghuo and {Lu}, Xing and {Sakai}, Takeshi},
        title = "{Digging into the Interior of Hot Cores with ALMA (DIHCA). II. Exploring the Inner Binary (Multiple) System Embedded in G335 MM1 ALMA1}",
      journal = {\apj},
         year = 2022,
        month = apr,
       volume = {929},
       number = {1},
          eid = {68},
        pages = {68},
          doi = {10.3847/1538-4357/ac5bd8},
archivePrefix = {arXiv},
       eprint = {2203.04333},
 primaryClass = {astro-ph.GA},
       adsurl = {https://ui.adsabs.harvard.edu/abs/2022ApJ...929...68O}
}

@ARTICLE{Olguin2023,
       author = {{Olguin}, Fernando A. and {Sanhueza}, Patricio and {Chen}, Huei-Ru Vivien and {Lu}, Xing and {Oya}, Yoko and {Zhang}, Qizhou and {Ginsburg}, Adam and {Taniguchi}, Kotomi and {Li}, Shanghuo and {Morii}, Kaho and {Sakai}, Takeshi and {Nakamura}, Fumitaka},
        title = "{Digging into the Interior of Hot Cores with ALMA: Spiral Accretion into the High-mass Protostellar Core G336.01-0.82}",
      journal = {\apjl},
         year = 2023,
        month = dec,
       volume = {959},
       number = {2},
          eid = {L31},
        pages = {L31},
          doi = {10.3847/2041-8213/ad1100},
archivePrefix = {arXiv},
       eprint = {2311.18006},
 primaryClass = {astro-ph.GA},
       adsurl = {https://ui.adsabs.harvard.edu/abs/2023ApJ...959L..31O}
}

@ARTICLE{QUARKSIII,
       author = {{Yang}, Dongting and {Liu}, Hong-Li and {Liu}, Tie and {Liu}, Xunchuan and {Xu}, Fengwei and {Qin}, Sheng-Li and {Tej}, Anandmayee and {Garay}, Guido and {Zhu}, Lei and {Mai}, Xiaofeng and {Jiao}, Wenyu and {Zhang}, Siju and {Dib}, Sami and {Stutz}, Amelia M. and {Palau}, Aina and {Sanhueza}, Patricio and {Zavagno}, Annie and {Yang}, A.~Y. and {Tang}, Xindi and {Tang}, Mengyao and {Zhang}, Yichen and {Garcia}, Pablo and {Zhang}, Tianwei and {Saha}, Anindya and {Li}, Shanghuo and {Goldsmith}, Paul F. and {Bronfman}, Leonardo and {Lee}, Chang Won and {Taniguchi}, Kotomi and {Ranjan Das}, Swagat and {Gorai}, Prasanta and {Hoque}, Ariful and {Chen}, Li and {Kou}, Zhiping and {Zhou}, Jianjun and {Zhang}, Yankun and {Toth}, L. Viktor and {Baug}, Tapas and {Shen}, Xianjin and {Li}, Chuanshou and {Zou}, Jiahang and {Das}, Ankan and {Nazeer}, Hafiz and {Dewangan}, L.~K. and {Hwang}, Jihye and {Chibueze}, James O.},
        title = "{The ALMA-QUARKS Survey: III. Clump-to-core fragmentation and search for high-mass starless cores}",
      journal = {arXiv e-prints},
         year = 2025,
        month = aug,
          eid = {arXiv:2508.03229},
        pages = {arXiv:2508.03229},
          doi = {10.48550/arXiv.2508.03229},
archivePrefix = {arXiv},
       eprint = {2508.03229},
 primaryClass = {astro-ph.GA},
       adsurl = {https://ui.adsabs.harvard.edu/abs/2025arXiv250803229Y}
}

@ARTICLE{RiveraSoto2020,
       author = {{Rivera-Soto}, Rudy and {Galv{\'a}n-Madrid}, Roberto and {Ginsburg}, Adam and {Kurtz}, Stan},
        title = "{Recombination Lines and Molecular Gas from Hypercompact H II regions in W51 A}",
      journal = {\apj},
         year = 2020,
        month = aug,
       volume = {899},
       number = {2},
          eid = {94},
        pages = {94},
          doi = {10.3847/1538-4357/aba749},
archivePrefix = {arXiv},
       eprint = {2007.12276},
 primaryClass = {astro-ph.GA},
       adsurl = {https://ui.adsabs.harvard.edu/abs/2020ApJ...899...94R}
}

@ARTICLE{Kim2001,
       author = {{Kim}, Kee-Tae and {Koo}, Bon-Chul},
        title = "{Radio Continuum and Recombination Line Study of Ultracompact H II Regions with Extended Envelopes}",
      journal = {\apj},
         year = 2001,
        month = mar,
       volume = {549},
       number = {2},
        pages = {979-996},
          doi = {10.1086/319447},
archivePrefix = {arXiv},
       eprint = {astro-ph/0010535},
 primaryClass = {astro-ph},
       adsurl = {https://ui.adsabs.harvard.edu/abs/2001ApJ...549..979K}
}

@ARTICLE{Kudritzki2000,
       author = {{Kudritzki}, Rolf-Peter and {Puls}, Joachim},
        title = "{Winds from Hot Stars}",
      journal = {\araa},
         year = 2000,
        month = jan,
       volume = {38},
        pages = {613-666},
          doi = {10.1146/annurev.astro.38.1.613},
       adsurl = {https://ui.adsabs.harvard.edu/abs/2000ARA&A..38..613K}
}

@ARTICLE{Vink2001,
       author = {{Vink}, Jorick S. and {de Koter}, A. and {Lamers}, H.~J.~G.~L.~M.},
        title = "{Mass-loss predictions for O and B stars as a function of metallicity}",
      journal = {\aap},
         year = 2001,
        month = apr,
       volume = {369},
        pages = {574-588},
          doi = {10.1051/0004-6361:20010127},
archivePrefix = {arXiv},
       eprint = {astro-ph/0101509},
 primaryClass = {astro-ph},
       adsurl = {https://ui.adsabs.harvard.edu/abs/2001A&A...369..574V}
}

@ARTICLE{Mottram2011,
       author = {{Mottram}, J.~C. and {Hoare}, M.~G. and {Urquhart}, J.~S. and {Lumsden}, S.~L. and {Oudmaijer}, R.~D. and {Robitaille}, T.~P. and {Moore}, T.~J.~T. and {Davies}, B. and {Stead}, J.},
        title = "{The Red MSX Source survey: the bolometric fluxes and luminosity distributions of young massive stars}",
      journal = {\aap},
         year = 2011,
        month = jan,
       volume = {525},
          eid = {A149},
        pages = {A149},
          doi = {10.1051/0004-6361/201014479},
archivePrefix = {arXiv},
       eprint = {1009.1774},
 primaryClass = {astro-ph.GA},
       adsurl = {https://ui.adsabs.harvard.edu/abs/2011A&A...525A.149M}
}

@ARTICLE{Kim2003,
       author = {{Kim}, Kee-Tae and {Koo}, Bon-Chul},
        title = "{Molecular Counterparts of Ultracompact H II Regions with Extended Envelopes}",
      journal = {\apj},
         year = 2003,
        month = oct,
       volume = {596},
       number = {1},
        pages = {362-382},
          doi = {10.1086/377579},
archivePrefix = {arXiv},
       eprint = {astro-ph/0306405},
 primaryClass = {astro-ph},
       adsurl = {https://ui.adsabs.harvard.edu/abs/2003ApJ...596..362K}
}

@ARTICLE{Garay1986,
       author = {{Garay}, Guido and {Rodriguez}, Luis F. and {van Gorkom}, J.~H.},
        title = "{Rotating and Expanding Ultracompact H II Regions}",
      journal = {\apj},
         year = 1986,
        month = oct,
       volume = {309},
        pages = {553},
          doi = {10.1086/164624},
       adsurl = {https://ui.adsabs.harvard.edu/abs/1986ApJ...309..553G}
}
\bibliographystyle{aasjournal}



\end{document}